\documentclass[12pt,aps,prd,english,superscriptaddress,eqsecnum,tightenlines,nofootinbib,showpacs]{revtex4}

\usepackage{colordvi}

\usepackage{color}

\usepackage[unicode=true,pdfusetitle,
bookmarks=true,bookmarksnumbered=false,bookmarksopen=false,
breaklinks=true,pdfborder={0 0 1},backref=false,colorlinks=false]
{hyperref}
\usepackage{upgreek}
\usepackage[T1]{fontenc}
\usepackage[utf8]{inputenc}
\usepackage{babel}
\usepackage{mathtools}
\usepackage{amsmath}
\usepackage{amssymb}
\usepackage{stmaryrd}
\usepackage{amsmath, bm}
\usepackage{bbm} 
\usepackage{amsfonts}
\usepackage{enumitem}
\usepackage{amsmath}
\usepackage{amssymb}
\usepackage{fancyhdr}
\usepackage{epsfig}
\usepackage{color}
\usepackage{amsfonts}
\usepackage{enumitem}
\usepackage{amsmath}
\usepackage{amssymb}
\usepackage{fancyhdr}
\usepackage{esint}
\usepackage{graphicx}
\usepackage{euscript,amssymb}
\usepackage{enumerate}
\usepackage{graphicx}
\usepackage{euscript,amssymb}
\usepackage{mathrsfs}
\usepackage{amsbsy}
\usepackage{epsfig}
\usepackage{amsfonts,amsthm}
\usepackage{amsbsy}
\usepackage{physics}

\begin{document}

\title{Fermions in a loop  quantum cosmological  spacetime}

\author{Yaser Tavakoli}
\email{yaser.tavakoli@guilan.ac.ir} 
\affiliation{Department of Physics, University of Guilan, Namjoo Boulevard, 41335-1914 Rasht, Iran}
\affiliation{Faculty of Physics, University of Warsaw, Pasteura 5, 02-093 Warsaw, Poland}

\author{Ahad Khaleghi Ardabili}
\email{akhaleghiardabili@pennstatehealth.psu.edu} 
\affiliation{Department of Anesthesiology and Perioperative Medicine, Penn State Milton S Hershey Medical Center, Hershey, Pennsylvania 17036, USA}
\affiliation{Critical Illness and Sepsis Research Center (CISRC), Penn State College of Medicine, Hershey, Pennsylvania 17036, USA}

\author{Sara Mosaddegh}
\email{mosaddegh.physics@gmail.com} 
\affiliation{Department of Physics, University of Guilan, Namjoo Boulevard, 41335-1914 Rasht, Iran}

\begin{abstract}

We present a detailed Hamiltonian treatment of an inhomogeneous  fermionic perturbation propagating on a closed Friedmann-Lema\^itre-Robertson-Walker spacetime quantized via loop quantum cosmology. Expanding the fermion in spinor harmonics on a spatial three-sphere and truncating at quadratic order, we derive  a decoupled, mode-by-mode Hamiltonian, where each mode behaves as a time-dependent Fermi oscillator. This framework naturally facilitates  a Schr\"odinger-picture evolution for fermionic perturbations on a quantum-corrected background.
Under the test-field approximation, each massive mode sees its own ``dressed metric,'' akin to bosonic perturbations, but with distinctive Planck-scale modifications in both temporal and spatial components. Massless modes, by contrast, experience only an equivalent class of conformally invariant backgrounds: quantum corrections drop out of the spatial sector, while the temporal component alone is dressed by quantum gravity corrections.
Extending beyond the test-field regime via a Born-Oppenheimer approximation, we incorporate fermionic backreaction self-consistently. Each mode's energy (depending on the either vacuum or pair states they occupied) sources a mode-dependent shift of the background Hamiltonian, resulting in a ``rainbow metric''. In the deep Planck regime, vacuum occupation yields a positive fermionic perturbation term in the Hamiltonian, raising the effective minisuperspace potential and delaying the bounce to higher density in the contracting branch. Excited occupation flips the perturbation sign, lowering the barrier and advancing the bounce to lower density and larger volume.  At large volumes, massive fermion backreaction settles into a constant energy density---an emergent cosmological constant---capable of driving late-time acceleration, whereas massless modes remain dynamically inert postbounce.
These effects break the exact time-reversal symmetry of the quantum bounce, modify the critical density condition, and generate asymmetric pre- and postbounce trajectories. Our results highlight the crucial role that fermionic degrees of freedom---through both their two-state spectrum and quantum-gravity couplings---play in shaping Planck-era and subsequent cosmological dynamics.

\end{abstract}


\date{\today}


\maketitle

\section{Introduction}

The study of quantum fields in the Planck regime necessitates a careful consideration of the quantization of the background spacetime on which these fields propagate. Loop quantum cosmology (LQC) \cite{Ashtekar:2003hd, Ashtekar:2011ni} emerges as a compelling framework for quantizing cosmological spacetimes at Planck scales, offering profound insights into the quantum origins of the Universe and its subsequent evolution. Rooted in the broader context of loop quantum gravity (LQG) \cite{Thiemann:2007pyv, Ashtekar:2004eh, Rovelli:2004tv}, LQC provides symmetry-reduced models that yield concrete results, such as resolving the big bang singularity by replacing it with a quantum bounce. Additionally, the theory predicts an upper bound on the energy density of the Universe, given by $\rho_{\rm cr} \sim 0.41\rho_{\rm Pl}$, where $\rho_{\rm Pl}$ denotes the Planck density \cite{Bojowald:2001xe, Ashtekar:2006wn, Ashtekar:2006rx, Ashtekar:2006uz}.

In LQC models coupled to a background matter source $T$ (e.g., a scalar field \cite{Ashtekar:2006rx} or dust  \cite{Husain:2011tk}), the matter field serves as a global relational time variable, facilitating the evolution of physical observables. Within this quantum geometric framework, the dynamics of quantum fields can be analyzed by examining the propagation of field modes on the quantized background. This approach has been explored for test scalar fields propagating on isotropic \cite{Ashtekar:2009mb} and anisotropic \cite{Dapor:2012jg} quantum backgrounds. Extensions of these studies to vector fields \cite{Lewandowski:2017cvz} and tensor perturbations \cite{Agullo:2012fc, Agullo:2012sh} have also been investigated. In such models, the Hamiltonians of scalar (or vector/tensor) fields can often be expressed as those of decoupled harmonic oscillators corresponding to free scalar modes. Consequently, the evolution of field modes in a quantum spacetime is equivalent to their evolution on an emergent ``dressed''  background metric, whose components depend on the quantum fluctuations of the original geometry. When backreaction effects are neglected, all modes perceive the same geometry \cite{Ashtekar:2009mb}. However, in the presence of backreactions, different modes may experience distinct geometries, influenced by the energies of the propagating modes \cite{Dapor:2012jg, Lewandowski:2017cvz, Parvizi:2021ekr}.

While many cosmological models predominantly focus on scalar fields as the primary matter source---justified by their dominant role in driving inflationary dynamics \cite{Mukhanov:2005sc}---a complete quantum gravitational description must incorporate all fundamental field types, including gauge fields \cite{Maleknejad:2012fw, Durrer:2013pga} and fermionic fields \cite{Ribas:2005vr, Ribas:2010zj}. Among these, spinorial degrees of freedom, such as Dirac fermions, are anticipated to play a foundational role in the matter content of the Universe within most fundamental theories. These fields could significantly influence early-Universe dynamics, potentially contributing to phenomena such as particle-antiparticle asymmetry \cite{Dolgov:1989us, Kuzmin:1998kk}, big bang nucleosynthesis \cite{Parker:1971pt}, or the formation of large-scale structures \cite{Chimento:2007fx, Cortez:2016xsn}. Additionally, fermionic fields may contribute to the Universe's accelerating expansion \cite{Ribas:2005vr, Enqvist:2012im}, underscoring their relevance in both early- and late-time cosmology. Investigating the behavior of Dirac fields on quantum cosmological backgrounds constitutes a critical step toward elucidating the role of matter in shaping the large-scale structure of the cosmos during the early epochs. The inclusion of fermionic fields alongside bosonic matter types is essential for achieving a comprehensive description of the Planck-era universe, where quantum gravitational effects dominate. Recent advancements have extended these investigations to LQC spacetimes \cite{ElizagaNavascues:2017adj, ElizagaNavascues:2018zzu, ElizagaNavascues:2019ccg, ElizagaNavascues:2019fkq, Bojowald:2007nu, Lewandowski:2021bkt, Scardua:2018omf}, providing new insights into the interplay between fermionic matter and quantum geometries.

In this paper, we aim to construct an LQC model that incorporates fermions while accounting for a single mode. Our approach extends previous studies that examined the  spectrum of  the gravitational field coupled to an inhomogeneous scalar field \cite{Ashtekar:2009mb} or a vector field \cite{Lewandowski:2017cvz}, up to the lowest nontrivial order in inhomogeneous perturbations. At this level of approximation, the inhomogeneous modes of bosonic fields (e.g., scalar or electromagnetic fields) decouple from the fermion modes, enabling us to analyze the fermion modes independently. This separation simplifies the analysis while preserving the essential physics of fermionic contributions to quantum cosmology.
Our primary goal is to investigate whether the theory of Dirac  field propagating on an LQC background can lead to the theory of the same field on a distinct dressed spacetime. Such dressed metrics would arise due to the interaction between fermionic matter and the underlying quantum geometry, potentially leading to novel phenomena such as mode-dependent rainbow metric scenarios \cite{Parvizi:2021ekr}. This exploration deepens our understanding of how quantum gravity influences early-Universe dynamics, revealing new insights into the interplay between fermionic matter and the quantum structure of spacetime. Ultimately, this work aims to bridge the gap between fundamental particle physics and quantum cosmology, advancing our understanding of the Universe's earliest moments.

This paper is organized as follows. In Sec.~\ref{sec:spinor-class}, we present a concise review of the theory of fermionic fields on a closed Friedmann-Lema\^itre-Robertson-Walker (FLRW)   background. This includes a detailed derivation of the Hamiltonian and the corresponding evolution equations governing the spinor modes, along with an examination of their Hilbert space structure. Furthermore, we analyze the spectrum of the fermion modes and their eigenstates, offering a physical interpretation of the results. 
In Sec.~\ref{sec:LQC}, we focus on the quantization of this cosmological background coupled to a massless scalar field $T$, which serves as a relational time parameter, within the framework of LQC. The quantum constraint in this setting leads to a Schr\"odinger-like equation describing the evolution of the background quantum geometry state with respect to the relational time. In Sec.~\ref{sec:Fermion-LQC}, by formulating the total Hamiltonian constraint for the coupled geometry-fermion system, we derive a time-dependent Schr\"odinger equation for each fermionic mode. Under the test-field approximation, where the backreaction of fermionic modes on the background spacetime is neglected, we obtain an evolution equation for the spinor modes on a classical-like closed FLRW background. This analysis reveals the emergence of an effective ``dressed'' metric governing the propagation of fermionic modes.
In Sec.~\ref{sec:BReffects}, we incorporate the backreaction of fermionic modes by employing a Born-Oppenheimer  approximation. We demonstrate that these backreaction effects lead to the emergence of a ``rainbow'' dressed metric, where different fermionic modes perceive distinct background geometries. This result highlights the intricate interplay between matter and geometry in the quantum gravitational regime.
A cosmological consequence of the backreaction effect is explored in Sec.~\ref{Sec:EarlyUniverse}, where its impact on early-Universe dynamics, including the timing of the quantum bounce and potential imprints on cosmic evolution, is analyzed.
Finally, in Sec.~\ref{sec:conclusion}, we summarize our key findings and discuss their broader implications for understanding the role of fermionic matter in quantum cosmology and the early Universe.

\section{Fermion  fields  in curved spacetime}
\label{sec:spinor-class}

This section follows the approach developed  in \cite{DEath:1986lxx} and outlines a Hamiltonian theory for fermionic perturbations around a homogeneous and isotropic minisuperspace background. The background is represented by a $k= + 1$ FLRW metric,  characterized by a single scale factor $a(x^0)$ and driven by a homogeneous massless scalar field  $T(x^0)$ (with $x^0$ being a generic temporal component). 
We expand the Dirac action using spinor harmonics defined on the spatial sections (three-spheres) of the spacetime. This expansion allows us to systematically derive the fermionic Hamiltonian expressed in terms of these  harmonics, providing a foundation to  analyze the quantum dynamics of the fermion modes in a cosmological context.

To investigate the behavior of the fermions  within an LQC background, we will employ a two-component spinor formalism, which is commonly utilized in studies related to quantum cosmology \cite{DEath:1984gmo,DEath:1986lxx}. Before proceeding, however, we wish to outline some key formulas that establish connections between this formalism and the more familiar framework of four-dimensional Dirac spinors (as discussed in Refs. \cite{Wess:1992cp, Barvinsky:1999yf})

\subsection{Two-component spinor formalism}
\label{sec:Two-componentF}

We employ two-component Weyl spinors $\{\phi_A, \chi_{A'}\}$ following the Van der Waerden notation. These spinors can be categorized into two distinct types, distinguished by either primed or unprimed  indices.  Their transformations under an element $M\in \mathrm{SL}(2, \mathbb{C})$,  are given by
\begin{align}
\tilde{\phi}_{A} &= M_A{}^B\, \xi_{B}, \quad\quad \quad 
\tilde{\phi}^{A} = (M^{-1})^A{}_B\,  \phi^{B},\nonumber \\
\tilde{\bar{\chi}}_{A'} &=
{M^{\ast}}_{A'}{}^{B'}\, \bar{\chi}_{B'},\quad \quad \tilde{\bar{\chi}}^{A'} =
{(M^{-1})^{*}}^{A'}{}_{B'}\, \bar{\chi}^{B'}.
\label{van-der}
\end{align}
Here,  we are using the index notation $A, B, \ldots=1,2$ and $A', B', \ldots=1',2'$ to denote the Grassmann variables \cite{Berezin:1976eg} forming the two-component spinors [see Eq.~(\ref{Dir-define}) below]. The spinors with primed indices transform under the  $\left(0,\frac12\right)$ representation of the Lorentz group, while those with unprimed  indices transform under the conjugate representation $\left(\frac12,0\right)$. Throughout this discussion, we adopt the Minkowski metric  $\eta_{ab}=\text{diag}(-1, 1, 1, 1)$.

The fundamental Pauli matrices, denoted as 
 $\sigma_{A A'}^{a}$ (with $a=0, 1, 2, 3$),  
 are defined as
\begin{align}
\sigma^0 = \left(
\begin{array}{rr}
-1\  &\  0   \\
0\  &\  -1
\end{array}\right), 
\quad 
\sigma^1 = \left(
\begin{array}{rr}
0\  &\  1   \\
1\  & \ 0
\end{array}
\right), 
\quad 
\sigma^2 = \left(
\begin{array}{rr}
0\  &\  -i   \\
i\  &\  0
\end{array}\right), 
\quad 
\sigma^3 = \left(
\begin{array}{rr}
1\  &\  0   \\
0\  &\  -1
\end{array}
\right).
\end{align}
Indices are raised and lowered using the alternating spinors $\epsilon^{AB}$,  $\epsilon^{A'B'}$ and $\epsilon_{AB}$, $\epsilon_{A'B'}$, each of which is given by the matrix
\begin{align}
\begin{pmatrix}
0 & 1\\
-1 & 0
\end{pmatrix}.
\label{eq:raiselowerM}
\end{align}
The analogous relations hold for the tensors with primed indices. Additionally, we define the conjugate Pauli matrices:
\begin{equation}
\bar{\sigma}^{a A'A} =
\epsilon^{A'B'} \epsilon^{AB}
\sigma^{a}_{B B'}.
\label{sigma-bar}
\end{equation}
These matrices satisfy the anticommutation relations
\begin{align}
&(\sigma^a \bar{\sigma}^b + \sigma^b\bar{\sigma}^a)_A^B
= -2\eta^{ab}\delta_A^B,\nonumber \\
&(\bar{\sigma}^a \sigma^b + \bar{\sigma}^b \sigma^a)_{B'}^{A'} = -2\eta^{ab}\delta_{B'}^{A'}.
\label{anticom}
\end{align}

For four-dimensional Dirac spinors, we adopt the Weyl basis, where the $\gamma$ matrices are represented as
\begin{equation}
\gamma^a = \left(
\begin{array}{ll}
0\  &\  \sigma^a \\
\bar{\sigma}^a\  &\  0
\end{array}
\right).
\label{gamma-define}
\end{equation}
In this basis, a Dirac spinor can be expressed in terms of the Weyl spinors
\begin{equation}
\Phi = \left(
\begin{array}{c}
\phi_A\vspace{1mm} \\
\bar{\chi}^{A'}
\end{array}
\right).
\label{Dir-define}
\end{equation}
Indices of these components are raised or lowered according to the matrix (\ref{eq:raiselowerM}); for instance,  $\phi_A=\phi^B\epsilon_{BA}$ and $\bar{\chi}^{A'}=\epsilon^{A'B'}\bar{\chi}_{B'}$.
The canonical representation of the Dirac $\gamma$ matrices,
\begin{equation}
\gamma_{\rm c}^0 = \left(
\begin{array}{rr}
-1\  &\  0 \\
0\  &\  1
\end{array}\right), \quad \quad  \
\gamma_{\rm c}^j = \left(
\begin{array}{cc}
0\  &\  \sigma^j \\
-\sigma^j\  &\  0
\end{array}\right),
\label{can-bas}
\end{equation}
is related to the Weyl basis by the similarity transformation
\begin{equation}
\Gamma_W = X \Gamma_{\rm c} X^{-1}, \quad \quad 
X = \frac{1}{\sqrt{2}}
\left(\begin{array}{rr}
1 & -1 \\
1 & 1
\end{array}\right).
\label{similarity}
\end{equation}
From a group-theoretical perspective, the Weyl basis is particularly advantageous because it naturally decomposes a Dirac spinor into irreducible representations of the Lorentz group [or its covering group $\mathrm{SL}(2,\mathbb{C})$].  Furthermore, this basis is especially useful in describing massless, ultrarelativistic fermions. Conversely, the canonical basis is often preferable for studying the nonrelativistic limit and deriving the Pauli equation.

Starting from the standard action for Dirac spinors in four-dimensional spacetime \cite{Itzykson:1980rh,Bogolyubov:1959bfo,Parker:2009uva,Birrell:1982ix} and employing the aforementioned relations, one obtains the action \cite{DEath:1984gmo}
\begin{align}
I_{\rm ferm} &= -\frac{i}{2}\int d^4x\, e \Big(\bar{\phi}^{A'}
e^{\mu}_{AA'}D_{\mu}\phi^A
+ \bar{\chi}^{A'}
e^{\mu}_{AA'}D_{\mu}\chi^A\Big)
+ \text{H.c.} \nonumber \\ 
&\quad\, - \frac{m}{\sqrt{2}}\int d^4x\, e \Big(\chi_A \phi^A
 + \bar{\phi}^{A'}\bar{\chi}_{A'}\Big).
\label{eq:actionFermion}
\end{align}
Here, $D_{\mu}$ is the spinorial covariant derivative, 
$e^{\mu}_{AA'}$ represents the contraction of the Pauli matrices with the tetrad $e^{\mu}_a$, and $e$ denotes the determinant of the tetrad. 
The corresponding Dirac equations are given by the variational condition, $\delta I_{\rm ferm} = 0$, with respect to fermion  fields,
\begin{subequations}
\label{eq:DiracFull}
\begin{align}
e_{AA'}^\mu D_\mu \phi^A &= i \frac{m}{\sqrt{2}} \bar{\chi}_{A'}, 
\label{eq:Dirac-eq1} \\
e_{AA'}^\mu D_\mu \chi^A &= i \frac{m}{\sqrt{2}} \bar{\phi}_{A'},
\label{eq:Dirac-eq2}\\
e_{AA'}^\mu D_\mu \bar{\phi}^{A'} &= -i \frac{m}{\sqrt{2}} \chi_A, 
\label{eq:Dirac-eq3}\\
e_{AA'}^\mu D_\mu \bar{\chi}^{A'} &= -i \frac{m}{\sqrt{2}} \phi_A.
\label{eq:Dirac-eq4}
\end{align}
\end{subequations}

\subsection{Harmonic expansion of fermions on  a closed FLRW background}
\label{sec:FermFLRW}

In this section, we study fermionic perturbations in the classical setting of a spatially closed  FLRW universe. The spacetime manifold is taken to be $M = \mathbb{S}^3 \times \mathbb{R}$, where spatial slices are modeled as three-sphere $\mathbb{S}^3$, and the temporal evolution is parametrized by $\mathbb{R}$. The line element in spherical coordinates $(\chi, \theta, \phi) \in \mathbb{S}^3$, with a temporal coordinate $x^0 \in \mathbb{R}$, is given by
\begin{equation}
ds^2 = -N_0(x^0) (dx^0)^2 + a^2(x^0) d\Omega_3^2,
\label{eq:metric}
\end{equation}
where $N_0(x^0)$ denotes the (homogeneous) lapse function, $a(x^0)$ is the scale factor, and $d\Omega_3^2$ represents the metric on the three-sphere $\mathbb{S}^3$,
\begin{equation}
d\Omega_3^2 = d\chi^2 + \sin^2\chi \left(d\theta^2 + \sin^2\theta d\phi^2\right).
\end{equation}
We assume the spatial sections are equipped with a fiducial Riemannian metric ${}^o q_{ij}$ corresponding to a round three-sphere of radius $a_o = 2$. The volume of the three-sphere with respect to this fiducial geometry is then given by $V_o = 2\pi^2 a_o^3 = 16\pi^2 \equiv \ell_o^3$, as commonly used in the LQC literature \cite{Szulc:2006ep}.  The physical volume of the spatial hypersurfaces $\mathbb{S}^3$ is then expressed as
\begin{equation}
V=\left(\frac{\ell_o}{2}\right)^3a^3\equiv\ell^3a^3.
\end{equation}
This setup provides a consistent classical background for studying the behavior of fermionic fields and sets the stage for quantization in both matter and geometric sectors.

To analyze fermionic perturbations on the background (\ref{eq:metric}), we expand the fermionic field using a complete set of spinor harmonics on the three-sphere $\mathbb{S}^3$.  These harmonics are eigenfunctions of the Dirac operators 
\begin{equation}
n_{AA'}e^{BA'j}\, {}^{(3)}D_j \quad \text{and} \quad n_{AA'}e^{AB'j}\, {}^{(3)}D_j,
\nonumber
\end{equation}
where $j=1,2,3$ denotes spatial indices,  ${}^{(3)}D_j$  represents the covariant derivative associated with the $\mathrm{SU(2)}$ spin connection on $\mathbb{S}^3$, and $n^{AA'}$ is the spinorial counterpart of the unit timelike normal to $\mathbb{S}^3$ \cite{DEath:1984gmo}. For  spinors with chirality $\phi_A$,  the basis is provided by the eigenmodes {\small{} \(\{\rho_A^{nq}, \bar{\sigma}_A^{nq}\}\)}, which satisfy the  eigenvalue equations
\begin{align}
-in_{AA'}e^{BA'j}\, {}^{(3)}D_j \rho_B^{nq}(\mathbf{x}) &=  + \lambda_n \rho_A^{nq}(\mathbf{x}),
\label{eq:eigenVSpin-rho1}\\
-in_{AA'} e^{BA'j}\, {}^{(3)}D_j \bar{\sigma}_{B}^{nq}(\mathbf{x}) &=  -\lambda_n\bar{\sigma}_{A}^{nq}(\mathbf{x}),
\label{eq:eigenVSpin-sigma2}
\end{align}
with eigenvalues $\lambda_n=n + \tfrac{3}{2}$, where $n=0, 1, 2, \ldots$ labels  each harmonic mode.   Each eigenvalue $\pm\lambda_n$  corresponds to a finite-dimensional eigenspace of dimension $d_n= (n+1)(n+2)$, reflecting the high symmetry of the three-sphere $\mathbb{S}^3$. This degeneracy  $d_n$ captures the number of linearly independent spinor harmonics associated with each energy level, indicating that every mode $n$ admits $d_n$ distinct fermionic states with the same eigenvalue. The additional label $q=1, 2, \ldots, d_n$ is introduced to distinguish these degenerate modes.
For spinors with opposite chirality, $\bar{\chi}^{A'}$, the basis is given by the Hermitian conjugates of the previous set, {\small{}$\{\bar{\rho}_{A'}^{nq}, \sigma_{A'}^{nq}\}$}, satisfying:
\begin{align}
-in_{AA'}e^{AB'j}\, {}^{(3)}D_j \bar{\rho}_{B'}^{nq}(\mathbf{x}) &=  -\lambda_n \bar{\rho}_{A'}^{nq}(\mathbf{x}), 
\label{eq:eigenVSpin-rho2}\\
-in_{AA'}e^{AB'j}\, {}^{(3)}D_j \sigma_{B'}^{nq}(\mathbf{x}) &=  + \lambda_n \sigma_{A'}^{nq}(\mathbf{x}).
\label{eq:eigenVSpin-sigma1}
\end{align}

The harmonics  {\small{} $\{\rho_A^{nq}, \bar{\sigma}_A^{nq}\}$} and  {\small{} $\{\bar{\rho}_{A'}^{nq}, \sigma_{A'}^{nq}\}$}  form orthogonal bases on $\mathbb{S}^3$,  satisfying the orthogonality relations
\begin{subequations}
\begin{align}
&\int d\mu \, \rho_A^{np}\, n^{AA'} \sigma_{A'}^{mq} = 0, \\
& \int d\mu \, \bar{\rho}_{A'}^{np}\, n^{AA'} \bar{\sigma}_A^{mq} = 0, \\
&\int d\mu \, \rho^{np}_A\, n^{AA'} \bar{\rho}_{A'}^{mq} = \delta^{nm} \delta^{pq}, \\
&\int d\mu \, \bar{\sigma}^{np}_A\, n^{AA'} \sigma_{A'}^{mq} = \delta^{nm} \delta^{pq}, 
\end{align}
\end{subequations}
where the integration measure on $\mathbb{S}^3$ is  $d\mu = \sin^2\chi \sin\theta \, d\chi \, d\theta \, d\phi$.
Using these eigenmodes, any two-component spinor field or its Hermitian conjugate on  $\mathbb{S}^3$ can be expanded as follows
\begin{subequations}
\label{eq:harmonicsEXP}
\begin{align}
\phi_A(x) &= \frac{a^{-\frac{3}{2}}}{2\pi} \sum_{npq} \alpha^{pq}_n \Big(m_{np}(x^0) \rho_A^{nq}(\mathbf{x}) + \bar{r}_{np}(x^0) \bar{\sigma}_A^{nq}(\mathbf{x})\Big), 
\label{eq:harmonics1}\\
\bar{\phi}_{A'}(x) &= \frac{a^{-\frac{3}{2}}}{2\pi} \sum_{npq} \alpha^{pq}_n \Big(\bar{m}_{np}(x^0) \bar{\rho}_{A'}^{nq}(\mathbf{x}) + r_{np}(x^0) \sigma_{A'}^{nq}(\mathbf{x})\Big),
\label{eq:harmonics2}\\
\chi_A(x) &= \frac{a^{-\frac{3}{2}}}{2\pi} \sum_{npq} \beta^{pq}_n \Big(s_{np}(x^0) \rho_A^{nq}(\mathbf{x}) + \bar{t}_{np}(x^0) \bar{\sigma}_A^{nq}(\mathbf{x})\Big),
\label{eq:harmonics3}\\
\bar{\chi}_{A'}(x) &= \frac{a^{-\frac{3}{2}}}{2\pi} \sum_{npq} \beta^{pq}_n \Big(\bar{s}_{np}(x^0) \bar{\rho}_{A'}^{nq}(\mathbf{x}) + t_{np}(x^0) \sigma_{A'}^{nq}(\mathbf{x})\Big).
 \label{eq:harmonics4}
\end{align}
\end{subequations}
In these expansions,  the summations are performed first over the degeneracy indices   $p$ and $q$ (each ranging from $1$ to $d_n$), followed by the mode number $n$  from $0$ to $\infty$. The functions  $m_{np}$, $r_{np}$, $s_{np}$, $t_{np}$,  and their complex conjugates encode the time evolution of the fermionic modes. These functions are treated as odd elements of a Grassmann algebra to ensure consistency with the anticommuting nature of fermionic fields. The coefficients \(\alpha^{pq}_n \) and \(\beta^{pq}_n\) simplify the structure of the action by eliminating couplings between different values of  $p$.  Explicitly, for a given $n$, they are block-diagonal matrices of dimension $d_n$,
\begin{equation}
\alpha^{pq}_n =
\begin{pmatrix}
1 & 1\\
1 & -1
\end{pmatrix} \quad \text{and} \quad 
\beta^{pq}_n =
\begin{pmatrix}
1 & -1\\
-1 & -1
\end{pmatrix}.
\end{equation}
The normalization factor $a^ {-3/2}$  ensures compatibility with the spatial volume of   $\mathbb{S}^3$, a standard feature in Hamiltonian formulations of fermionic fields in globally hyperbolic spacetimes   \cite{Nelson:1978ex}. This scaling arises naturally when deriving Dirac brackets after eliminating second-class constraints \cite{Dirac:1964}.

This formalism provides a rigorous framework for studying the Hamiltonian dynamics of fermionic perturbations in a closed Friedmann universe. By decomposing the fields into eigenmodes and incorporating their time evolution, it enables a detailed analysis of their classical and quantum effects on cosmological backgrounds.

\subsection{Hamiltonian formalism of fermions}

By substituting the fermionic field expansion (\ref{eq:harmonicsEXP}) into the action functional (\ref{eq:actionFermion}) and utilizing the properties of spinor harmonics, we obtain the following form for the action:
\begin{align}
I_{\rm ferm} &= \sum_{np}\int dt N_0\Big[\frac{i}{2N_0}
\big(\bar{m}_{np}\dot{m}_{np} + m_{np}\dot{\bar{m}}_{np}
+ \bar{s}_{np}\dot{s}_{np} + s_{np}\dot{\bar{s}}_{np}
\nonumber \\
&\qquad\qquad \qquad \qquad \quad + \bar{t}_{np}\dot{t}_{np} + t_{np}\dot{\bar{t}}_{np}
+ \bar{r}_{np}\dot{r}_{np} + r_{np}\dot{\bar{r}}_{np}\big)
\nonumber \\
&\qquad\qquad \qquad   + \lambda_na^{-1}\big(\bar{m}_{np}m_{np}
+ \bar{s}_{np}s_{np} + \bar{t}_{np}t_{np}
+ \bar{r}_{np}r_{np}\big) \nonumber \\
&\qquad\qquad \qquad   - m(r_{np}t_{np} + \bar{t}_{np}\bar{r}_{np} + s_{np}m_{np} + \bar{m}_{np}\bar{s}_{np})\Big].             
\label{action1}
\end{align}
For each mode $(n,p)$, rewriting the action  in a more compact form, it takes the structure
\begin{align}
I_{np} &= \int dt N_0 \left(\frac{i}{2N_0}(\bar{x}\dot{x}
+x\dot{\bar{x}} + \bar{y}\dot{y}
+y\dot{\bar{y}})+\frac{\lambda_n}{a}(\bar{x}x
+ \bar{y}y) - m(yx + \bar{x}\bar{y})\right),                
\label{eq:action2}
\end{align}
where, $(x, y)$ denote the Grassmann variables associated with the fermionic degrees of freedom, $(m_{np}, s_{np})$ or $(t_{np}, r_{np})$. The total fermion action may thus be expressed as 
\begin{align}
I_{\rm ferm} &= \sum_{np}\big[I_{np}(m_{np}, \bar{m}_{np}, s_{np}, \bar{s}_{np}) + I_{np}(r_{np}, \bar{r}_{np}, t_{np}, \bar{t}_{np})\big]  \nonumber \\
&=: \sum_{np} I_{np}(x, y, \bar{x}, \bar{y}).
\end{align}
The action (\ref{eq:action2}) leads to the equations of motion for the fermionic field components which are given by
\begin{subequations}
\begin{align}
& i\frac{\dot{x}}{N_0} + \nu x - m\bar{y} = 0,
\label{eq:Dirac-eq1b}\\
& i\frac{\dot{\bar{x}}}{N_0} - \nu \bar{x} + my = 0,
\label{eq:Dirac-eq2b}\\
& i\frac{\dot{y}}{N_0} + \nu y + m\bar{x} = 0,
\label{eq:Dirac-eq3b}
\\
&i\frac{\dot{\bar{y}}}{N_0} - \nu \bar{y} - mx = 0,
\label{eq:Dirac-eq4b}
\end{align}
\end{subequations}
where 
\begin{equation}
\nu(x^0)= a^{-1}(x^0)(n + \tfrac{3}{2})  \equiv\lambda_na^{-1}(x^0).
\end{equation}  
These equations are simply the Dirac equations expressed in terms of harmonic modes $(x,y)$. However, it is often useful to recast them into a second-order differential form. By doing so, we find that the variables $x$ and $y$ satisfy an identical second-order equation given by
\begin{align}
\frac{1}{N_0}\frac{d}{dt}\left(\frac{\dot{x}}{N_0}\right) + \left(\frac{\dot{\nu}}{iN_0} + \nu^2 +m^2\right)x=0.
\end{align}
Similarly, the conjugate variables $\bar{x}, \bar{y}$ obey the corresponding equation
\begin{align}
\frac{1}{N_0}\frac{d}{dt}\left(\frac{\dot{\bar{x}}}{N_0}\right) + \left(-\frac{\dot{\nu}}{iN_0} + \nu^2 +m^2\right)\bar{x}=0.
\end{align}
These second-order formulations provide a more direct way to analyze the evolution of the fermionic perturbations within the given background.

Following the Dirac procedure, one may obtain the
Hamiltonian
\begin{equation}
H_{np}(x, \bar{x}, y, \bar{y}) = N_0 \left[\nu(x\bar{x} + y\bar{y})+m(yx+ \bar{x} \bar{y})\right],
\label{eq:HamiltonianFerm1}
\end{equation}
where $x, \bar{x}, y, \bar{y}$ satisfy the fundamental Dirac bracket relations 
\begin{equation}
[x, \bar{x}]^\ast =-i \quad \text{and} \quad [y, \bar{y}]^\ast =-i,
\label{eq:DiracBracket1}
\end{equation}
with all other brackets vanishing.  The total Hamiltonian describing the dynamics of the fermionic perturbations is then obtained by summing over all allowed $n,p$, yielding
\begin{align}
H_{\rm ferm} &=  \sum_{np}\big[H_{np}(m_{np}, \bar{m}_{np}, s_{np}, \bar{s}_{np}) + H_{np}(r_{np}, \bar{r}_{np}, t_{np}, \bar{t}_{np})\big] \nonumber \\
&=: \sum_{np}H_{np}(x, \bar{x}, y, \bar{y}).
\label{eq:HamiltonianEq.Tot}
\end{align}
This  provides a canonical description of the fermionic sector, incorporating the effects of the evolving background geometry through the time-dependent parameters $\nu(x^0)$.

We have reformulated the problem by replacing the original fermionic field variables  $\{\phi_A, \chi_A, \bar{\phi}_{A'}, \bar{\chi}_{A'}\}$ with the expansion functions  $\{m_{np}, s_{np}, t_{np}, r_{np}\}$ and their conjugate quantities, as defined in Eqs.~(\ref{eq:harmonicsEXP}).  At the lowest nontrivial order in perturbation theory, these coefficients remain invariant under local Lorentz transformations and diffeomorphisms on $\mathbb{S}^3$. Consequently, the generators of these transformations do not contribute to the dynamics of this model. Therefore, for the order of approximation considered here, it suffices to impose only the  constraint given by the  sum of the background Hamiltonian and the  fermion Hamiltonian (\ref{eq:HamiltonianEq.Tot}),
\begin{equation}
H_{\rm geo} + H_{\rm ferm} \approx 0.
\label{eq:Hgeo-0}
\end{equation}
Here, $H_{\rm geo}=H_{\rm grav}+H_T$   combines the gravitational Hamiltonian $H_{\rm grav}$ with the homogeneous scalar‐field contribution $H_T$.

In the quantum theory, the classical fermionic Hamiltonian (\ref{eq:HamiltonianFerm1}) is promoted to an operator by quantizing the underlying dynamical variables in accordance with their anticommutation properties. Specifically, the fundamental anticommutation relations are imposed as
\begin{equation}
\{\hat{x}, \hat{\bar{x}}\} =1 \quad \text{and} \quad \{\hat{y}, \hat{\bar{y}}\}=1,
\label{eq:AntiCommutation}
\end{equation}
with all other anticommutators vanishing. These relations arise naturally from the Dirac bracket structure (\ref{eq:DiracBracket1}) through canonical quantization rules for fermionic (Grassmann-valued) degrees of freedom \cite{Casalbuoni:1975bj}. A convenient realization of these anticommutation relations is provided by the ``holomorphic representation'' \cite{Berezin:1966nc, Faddeev:1980be}, in which the conjugate variables are represented as functional derivatives, 
\begin{equation}
\bar{x} \to \hat{\bar{x}}=\partial/\partial x \quad \text{and} \quad \bar{y} \to \hat{\bar{y}}=\partial/\partial y,
\end{equation}
In this representation, the quantum state is described by a wave function $\psi$  that depends only on the unbarred Grassmann variables,  $\psi(x, y)\equiv\psi(m, s, r, t)$,  for each harmonic mode labeled by $(n,p)$. The dynamics of these quantum fermionic modes are governed by the  Schr\"odinger equation, 
\begin{align}
i\hbar\partial_{x^0} \psi(x^0; x,y) &= \sum_{np} \hat{H}_{np}(x,y)\psi(x^0; x,y),
\label{eq:Schrodinger-Class0}
\end{align}
where the mode Hamiltonians take the form
\begin{equation}
\hat{H}_{np} = N_0 \left[\nu(\widehat{x\bar{x}} + \widehat{y\bar{y}})+m(\widehat{yx}+ \widehat{\bar{x}\bar{y}})\right].
\label{eq:Schrodinger-Class1-A}
\end{equation}
As will be discussed in Secs.~\ref{subsec:ClosedFLRWScalarF} and \ref{sec:Fermion-LQC}, we  adopt a harmonic time gauge, in which the lapse is chosen as $N_\tau=N(\tau)=a^3$, simplifying the coupling between geometry and matter. In this case, the Hamiltonian (\ref{eq:Schrodinger-Class1-A}) can be reexpressed as
\begin{align}
\hat{H}_{np}^{(\tau)} &= \lambda_n a^2(\widehat{x\bar{x}} + \widehat{y\bar{y}})+ma^3(\widehat{yx}+ \widehat{\bar{x}\bar{y}}) 
\nonumber \\
&= \lambda_n \ell^{-2}V^{2/3}(\widehat{x\bar{x}} + \widehat{y\bar{y}})+m\ell^{-3}V(\widehat{yx}+ \widehat{\bar{x}\bar{y}}).
\label{eq:Schrodinger-Class1}
\end{align}

Since the perturbation modes are dynamically decoupled, the total wave function $\psi(x, y)$ naturally factorizes into a product over individual modes,
\begin{equation}
\psi(x,p) = \prod_{np}\psi_{np}(m_{np}, s_{np}, r_{np}, t_{np}),
\label{eq:wavefucntion-Fmodes}
\end{equation}
where $\psi_{np}$ depends solely on the Grassmann variables associated with the $(n,p)$th mode. This factorized ansatz ensures that the action of  $\hat{H}_{np}^{(\tau)}$  is restricted to its corresponding mode, leaving all other components of the wave function unaffected. This reflects the physical independence and linear superposition of fermionic perturbations in the linearized regime.

The inner product for (each mode's) wave functions  $\psi_1(x, y)$ and $\psi_2(x, y)$,  in the holomorphic representation is defined as
\begin{equation}
\langle\psi_1, \psi_2\rangle = \int \overline{\psi_1(x, y)}\, \psi_2(x,y) e^{-x\bar{x} - y\bar{y}}\, dx\, d\bar{x}\, dy\, d\bar{y}.
\label{eq:InnerProduct}
\end{equation}
This inner product ensures that the Hamiltonian operator (\ref{eq:Schrodinger-Class1}) for each mode $(n,p)$, is self-adjoint, thereby defining the structure of the Hilbert space $\mathscr{H}_{np}$. The exponential factor
$e^{-x\bar{x} - y\bar{y}}$ ensures convergence of the integral, while integration is performed according to the Berezin rules \cite{Berezin:1966nc, Faddeev:1980be}, which consistently handle Grassmann variables ($x$, $y$, $\bar{x}$, $\bar{y}$) and their derivatives. Specifically, the Berezin integration rules are \cite{Berezin:1966nc, Faddeev:1980be}
\begin{equation}
\int dx=0, \quad \int xdx=1, \quad \int d\bar{x}=0, \quad \int \bar{x}d\bar{x}=1,
\end{equation}
with analogous rules applying to $y$ and $\bar{y}$. 

Stationary states   $\psi_{np}(x, y)$ of the Hamiltonian (\ref{eq:Schrodinger-Class1}) satisfy the eigenvalue equation,
\begin{equation}
\hat{H}_{np}^{(\tau)}\psi_{np}=E_{np}\psi_{np},
\end{equation}
and are orthogonal with respect to the inner product (\ref{eq:InnerProduct}). This orthogonality ensures that the eigenstates form a complete basis for the Hilbert space $\mathscr{H}_{np}$, associated with each mode $(n,p)$.
Once the mode-dependent subspaces $\mathscr{H}_{np}$ are constructed using the eigenstates   $\psi_{np}$, the total Hilbert space for the fermionic perturbations is obtained as the direct product of these subspaces, 
\begin{equation}
\mathscr{H}_{\rm ferm} = \bigotimes_{np} \mathscr{H}_{np}.
\end{equation}
This direct product structure reflects the decoupling of the perturbation modes, allowing each mode to evolve independently within its corresponding subspace.

\subsection{Spectral analysis of fermionic modes}

In the transition from the classical constraint \eqref{eq:Hgeo-0} to its  quantum version, an inherent operator-ordering ambiguity arises.  Indeed, terms involving products such as $x\bar{x}$ require careful treatment, as they involve noncommuting Grassmann variables.

To address this, we adopt the Weyl symmetric ordering \cite{Berezin:1976eg, Henneaux:1982ma}, which symmetrizes products of operators to yield a consistent and Hermitian quantization scheme. Specifically, we replace products like $x\bar{x}$ with their symmetrized counterpart,
\begin{equation}
x\bar{x} \to \frac{1}{2}\left(x\frac{\partial}{\partial x} - \frac{\partial}{\partial x}x\right).
\end{equation}
Under this prescription, for each  fermionic mode, the Hamiltonian \eqref{eq:Schrodinger-Class1} on the classical gravitational background takes the explicit form
\begin{align}
\hat{H}_{np}^{(\tau)} &= \lambda_n\ell^{-2}V^{\frac{2}{3}} (\widehat{x\bar{x}} + \widehat{y\bar{y}})+m \ell^{-3}V(\widehat{yx}+ \widehat{\bar{x}\bar{y}}) \nonumber \\
&=  \lambda_n\ell^{-2}  V^{\frac{2}{3}}\left(-1+x\frac{\partial}{\partial x} + y\frac{\partial}{\partial y}\right) + m\ell^{-3} V \left(yx+ \frac{\partial^2}{\partial x\partial y}\right).
\label{eq:Eigenvalue1in}
\end{align}
The corresponding eigenvalue equation is then
\begin{align}
&\left[\frac{\lambda_n}{\ell^2}V^{2/3}\left(-1+x\frac{\partial}{\partial x} + y\frac{\partial}{\partial y}\right) + \frac{m}{\ell^3}V\left(yx+ \frac{\partial^2}{\partial x\partial y}\right)\right] \psi_{np}(x,y)
= E_{np}\psi_{np}(x,y), 
\label{eq:Eigenvalue1}
\end{align}
governing the quantum states $\psi_{np}(x, y)$. The full Hamiltonian \eqref{eq:HamiltonianEq.Tot} contains two such contributions per mode---one for each helicity state---ensuring the proper accounting of spinor degrees of freedom.

To solve the eigenvalue equation~\eqref{eq:Eigenvalue1}, we expand the wave function in terms of the Grassmann variables  $x$, $y$. This expansion naturally yields a four-dimensional Hilbert space, spanned by a basis constructed from the monomials $1$, $x$, $y$, and $xy$ (cf. Appendix~\ref{sec:eigenvalue}). The resulting eigenstates and corresponding eigenvalues are given by
\begin{subequations}
\label{eq:EigenValueSoltogenTOT}
\begin{align}
&E_{np}^{(0)}=-w_{n}: \quad 
\psi_{np}^{(0)} = \mathrm{N}_{n}^{(0)}\Big(1+\frac{m\ell^{-3}V}{\lambda_n\ell^{-2} V^{2/3} + w_{n}}xy\Big),     \label{eq:EigenValueSoltogen-s1} \\
&E_{np}^{(1)}=0: \ \, \quad \quad \psi_{np}^{(1)} = \mathrm{N}_{n}^{(1)}x,    
\label{eq:EigenValueSoltogen-s2}\\
& E_{np}^{(2)}=0: \ \, \quad\quad   \psi_{np}^{(2)} = \mathrm{N}_{n}^{(2)}y,  
\label{eq:EigenValueSoltogen-s3}\\
& E_{np}^{(3)}=+w_{n}: \quad  \psi_{np}^{(3)} = \mathrm{N}_{n}^{(3)}\Big(1+\frac{m\ell^{-3}V}{\lambda_n\ell^{-2}V^{2/3} -w_{n}}xy\Big).
\label{eq:EigenValueSoltogen-s4}
\end{align}
\label{eq:EigenValueSoltogen}
\end{subequations}
The quantity $w_n$ defines the characteristic energy scale of the system and is given by
\begin{equation}
w_n=\sqrt{\frac{\lambda_n^2}{\ell^4}V^{\frac{4}{3}}+\frac{m^2}{\ell^6}V^{2}}.
\label{eq:omega-energy}
\end{equation}
The normalization constants  $\mathrm{N}_{n}^{(I)}$ (with $I=0, \ldots, 3$) are determined using the fermionic inner product defined in Eq.~\eqref{eq:InnerProduct} and are found to be 
\begin{align}
\mathrm{N}_{n}^{(0)}=\Big(\frac{\lambda_n\ell^{-2} V^{2/3} +w_{n}}{2w_{n}}\Big)^{\frac{1}{2}}, \quad \quad \mathrm{N}_{n}^{(1)}=\mathrm{N}_{n}^{(2)}=1,\quad \quad  \mathrm{N}_{n}^{(3)}=\Big(\frac{\lambda_n\ell^{-2} V^{2/3}  -w_{n}}{2w_{n}}\Big)^{\frac{1}{2}}.
\end{align}
Although the wave functions $\psi_{np}^{(I)}(x, y)$ carry both mode labels  $n$ and $p$, the corresponding energy eigenvalues and normalization constants depend only on the principal label $n$. This reflects the fact that, for a given mode number $n$, the energy spectrum is independent of the helicity label $p$. Consequently, we omit the $p$ dependence when referring to energy levels.

The general state for a given mode $(n, p)$ is a linear combination \begin{equation} 
\psi_{np}(x, y) = \sum_{I=0}^3 a_I \psi_{np}^{(I)}(x, y), \end{equation}
with complex coefficients $a_J$. The Grassmann structure ensures that the fermionic Hilbert space is ``finite-dimensional,'' as expected from the anticommutation relations \eqref{eq:AntiCommutation}. The resulting  bases [listed in Eq.~\eqref{eq:EigenValueSoltogen}] correspond to the occupation states of fermionic creation and annihilation operators for each mode $(n,p)$.
Physically, the four states $\psi_{np}^{(I)}(x, y)$ describe the vacuum, single-particle, single-antiparticle, and particle-antiparticle configurations: 
\begin{itemize}
\item[i)] The ground state $\psi_n^{(0)}$ with $E_n^{(0)} = -w_n$ represents the Dirac vacuum.
\item[ii)] The degenerate zero-energy states $\psi_n^{(1)}$ and $\psi_n^{(2)}$ describe single-antiparticle and single-particle excitations, respectively.
\item[iii)] The state $\psi_n^{(3)}$ with $E_n^{(3)} = +w_n$ corresponds to a particle-antiparticle pair.
\end{itemize}
This structure manifests key features: vacuum energy, degenerate single-excitation states, and symmetric particle-antiparticle behavior.

In the massless limit ($m=0$), the $m$-dependent mixing term  in Eq.~(\ref{eq:Eigenvalue1in}) drops out. The Hamiltonian then simplifies to
\begin{align}
\hat{H}_{np}^{(\tau)} &= \lambda_n\ell^{-2}V^{\frac{2}{3}} (\widehat{x\bar{x}} + \widehat{y\bar{y}}),
\label{eq:Eigenvalue1in-mzero}
\end{align}
which is manifestly diagonal in the occupation-number basis  $\{1, x, y, xy\}$. In particular, there is no coupling between the vacuum state  $\psi_n^{(0)}$  and the two-particle state $\psi_n^{(3)}$, so particle-antiparticle pair creation is forbidden.
Moreover, in a conformally flat FLRW background, massless fermions respect the spacetime's conformal symmetry. Consequently, their mode functions can be conformally mapped to those in Minkowski space---where, by construction, no particle production takes place.

\section{Quantized closed FLRW background}
\label{sec:LQC}

In this section, we present a concise review of the quantization framework for the background closed FLRW geometry within the context of LQC. Based on this formalism, a Schr\"odinger-like equation is derived that describes the evolution of the background quantum state. This derivation provides the essential mathematical foundation for investigating the dynamics of the quantum universe and perturbations in the Planck regime.

\subsection{Background coupled to a scalar  field}
\label{subsec:ClosedFLRWScalarF}

We focus on the closed FLRW models (\ref{eq:metric}), coupled with a massless scalar field $T(x^0)$, serving as the background matter source. The dynamics of the system are governed by the Hamiltonian constraint, expressed as $N_0H_{\rm geo}\approx0$, where $H_{\rm geo}$ represents the Hamiltonian of the background geometry. This constraint takes the form
\begin{equation}
N_0H_{\rm geo} = N_0(H_T + H_{\rm grav})  \approx 0,
\label{eq:geometryHConstraint}
\end{equation}
with the weak equality holding on the constraint hypersurface.  Here, $H_T$ and $H_{\rm grav}$ denote the Hamiltonians for the scalar field and the gravitational sector, respectively.

The Hamiltonian for the gravitational sector in the closed FLRW model is given by  \cite{Ashtekar:2006es}
\begin{align}
H_{\rm grav} &= -\frac{3}{8\pi G}\frac{\sqrt{p}}{\gamma^2}
\left[ \left(c-  \frac{\ell_o}{a_o} \right)^2 - \frac{\gamma^2\ell_o^2}{a_o^2}\right].
\end{align}
The connection and triad variables are defined as $c=\gamma\dot{a}$ and $|p|=a^2\ell_o^2/4=a^2\ell^2$, respectively, where $a$ is the scale factor associated with the physical spatial metric $q_{ij}$. At each phase space point $(c, p)$, the spatial metric is expressed as $q_{ij}={}^oq_{ij}\, |p|\ell_o^{-2}$. These variables satisfy the fundamental Poisson bracket
\begin{equation}
\{c,p\}=\frac{8\pi G\gamma}{3},
\end{equation}
with  $\gamma$ denoting the Barbero-Immirzi parameter of LQG. The corresponding physical volume of  $\mathbb{S}^3$, in terms of triad variable $p$,  is given by
\begin{equation}
V=\ell^3a^3=|p|^{3/2}.
\end{equation}
The matter content of the model is described by a massless scalar field $T$, with conjugate momentum $P_T$, forming the canonical pair  $(T, P_T)$. The Hamiltonian for this homogeneous scalar field is
\begin{equation}
H_T = \frac{P_T^2}{2V},
\end{equation}
where $P_T$ is a positive constant, reflecting the free evolution of the massless field. 

To analyze the dynamics of inhomogeneous test fermion fields on this background near the Planck regime (as developed in Sec.~\ref{sec:Fermion-LQC}), we adopt the harmonic time gauge, defined by the condition  $\square \tau = 0$. This choice corresponds to a lapse function  $N_\tau=a^3$, which brings significant simplification to the scalar constraint and facilitates a natural deparametrization of the system using the homogeneous massless scalar field $T$ as a  relational clock.  
In solutions to the field equations $\square T=0$, the scalar field $T$ grows linearly in the harmonic time $\tau$ as $T = (P_T/\ell^3)\tau$.  While $T$ qualifies it as a relational time  in LQC,  this allows interchangeable use of $\tau$ and $T$. Under this gauge choice, the  constraint (\ref{eq:geometryHConstraint})  takes the following form: 
\begin{align}
H_{{\rm geo}}^{(\tau)} := N_\tau H_{\rm geo} &= \frac{P_T^2}{2\ell^3}  -\frac{3}{8\pi G}\frac{1}{\gamma^2\ell^3}
V^{2/3}\left[ \left(c-  \frac{\ell_o}{2}\frac{2}{a_o} \right)^2 - \frac{\gamma^2\ell_o^2}{4}\frac{4}{a_o^2}\right] V^{2/3} \nonumber \\
&=:  \frac{P_T^2}{2\ell^3}  + H_{{\rm grav}}^{(\tau)}
\approx 0.
\label{eq:HamiltonianLQC-Class}
\end{align}
Although the harmonic time coordinate may become globally ill defined in global extensions of closed FLRW spacetimes, it remains regular and physically meaningful within any finite temporal interval surrounding the quantum bounce, whether during the contracting or expanding phase. In this Planck-scale regime, the dynamics are governed by LQC modifications that ensure a smooth evolution of the background geometry. Consequently, the harmonic time gauge offers a robust and physically consistent framework for analyzing the truncated dynamics and formulating the zero-mode scalar constraint \cite{Ashtekar:2009mb}.

Within the test-field approximation---where the backreaction of the inhomogeneous fermionic field (characterized by the mode functions $m_{np}$, $r_{np}$, $s_{np}$, $t_{np}$ and their conjugates)  on the quantum geometry is neglected---the infinitely many  phase space constraints reduce to the zero-mode scalar constraint. Only the Hamiltonian constraint smeared with homogeneous lapses survives; Gauss and diffeomorphism constraints become trivial. The harmonic gauge thus combines technical simplicity with physical clarity for truncated system dynamics.

\subsection{Background quantization in LQC}

To quantize the background spacetime coupled to a scalar field, described by the variables $V$ and $T$, we adopt the LQC framework. In this approach, the scalar field  $T$ is treated using the standard Schr\"odinger representation, while the volume variable $V$ is quantized using the polymer representation to correctly incorporate quantum geometric effects. 

The quantum counterpart of the Hamiltonian constraint (\ref{eq:HamiltonianLQC-Class}) can be written as a self-adjoint operator on the kinematical Hilbert space $\mathscr{H}_{\rm kin}^o$,
\begin{equation}
\hat{H}_{{\rm geo}}^{(\tau)} \Psi_o(v, T) = \left(\frac{\hat{P}_T^2}{2\ell^3}  + \hat{H}_{\rm grav}^{(\tau)} \right)\Psi_o(v, T) \approx 0.
\end{equation}
Here, the quantized gravitational Hamiltonian for the closed FLRW model reads \cite{Ashtekar:2006es} 
\begin{align}
\hat{H}_{\rm grav}^{(\tau)} &= \frac{1}{16\pi G\ell^3}\hat{V}^{\frac{1}{2}}e^{i f \ell_o} \sin(\bar{\mu} c) \hat{A} \sin(\bar{\mu} c) e^{-i f \ell_o}\hat{V}^{\frac{1}{2}} \nonumber \\
& \quad -   \frac{1}{16\pi G\ell^3}\hat{V}^{\frac{1}{2}}\left( \sin^2 \left( \frac{\bar{\mu}\ell_o}{2} \right) - \frac{\bar{\mu}^2\ell_o^2}{4} -\frac{\ell_o^2}{9 |K^2 v|^{2/3}} \right) \hat{A}\hat{V}^{\frac{1}{2}},
\label{eq:Hamiltonian-GravQ}
\end{align}
where \(\bar{\mu}=\sqrt{\Delta}|p|^{-1/2}\)  is a specific function of $p$ which is related to the area gap $\Delta\equiv (2\sqrt{3}\pi\gamma) \ell_{\rm Pl}^2$  in LQC \cite{Ashtekar:2006wn}.  The function \( f(v) \) is  continuous  in $v$, ensuring the proper quantum representation. The parameter
\( v \) is a discrete (dimensionless) label related to the eigenvalues of the volume operator $\hat{V}=|\hat{p}|^{3/2}$  in LQC \cite{Ashtekar:2006wn},
\begin{equation}
\hat{V}|v\rangle = \left(\frac{8\pi \gamma}{6}\right)^{3/2} \frac{|v|}{K} \ell_{\rm Pl}^3 |v\rangle,
\label{eq:volumeoperatorEV}
\end{equation}
where $K=2\sqrt{2}/(3\sqrt{3\sqrt{3}})$.
Furthermore, \( \hat{A} (v) \) is an operator given by
\begin{align}
\hat{A} \Psi_o(v) &= \frac{24i \text{sgn}(\mu)}{8\pi\gamma^3\bar{\mu}^3\ell_{\rm Pl}^2}\left[\sin\left(\frac{\bar{\mu}c}{2}\right)\hat{V}\cos\left(\frac{\bar{\mu}c}{2}\right) - \cos\left(\frac{\bar{\mu}c}{2}\right) \hat{V}\sin\left(\frac{\bar{\mu}c}{2}\right)\right]\Psi_o(v)
  \nonumber \\
  &= -\frac{27K}{4} \sqrt{\frac{8\pi}{6}} \frac{\ell_{\rm Pl}}{\gamma^{3/2}} |v| \big||v-1| - |v+1| \big| \Psi_o(v).
 \label{eq:A-hat}
\end{align}

The  Hamiltonian (\ref{eq:Hamiltonian-GravQ})  governs the quantum evolution of the gravitational sector in the closed \( k=1 \) universe within LQC. In addition, on the matter Hilbert space $L^2(\mathbb{R}, dT)$, the dynamical variables $T$ and
its conjugate momentum $P_T$ are promoted to operators on it according to the Schr\"odinger picture: $\hat{T}=T$ and $\hat{P}_T=-i\hbar(\partial/\partial T)$. 
The total Hamiltonian constraint (\ref{eq:Hamiltonian-GravQ}) of the geometry $\hat{H}_{{\rm geo}}^{(\tau)}$ is written  now as
\begin{align}
\hat{H}_{{\rm geo}}^{(\tau)}\Psi_o(v, T)  &= -\frac{\hbar^2}{2\ell^3}\left(\partial_T^2 - \frac{2\ell^3}{\hbar^2}\hat{H}_{\rm grav}^{(\tau)}\right)\Psi_o(v, T) = 0.
\end{align}
This simplifies to the following equation for the wave function $\Psi_o(v, T)$
\begin{align}
\partial_T^2\Psi_o(v, T)  &=   -\Theta\Psi_o(v, T),
\end{align}
where, $\Theta$ is a difference operator that acts on $\Psi_o(v)\in\mathscr{H}_{\rm kin}^{\rm grav}$, defined as:
\begin{align}
 \Theta\Psi_o(v) &= -\frac{2\ell^3}{\hbar^2}\hat{H}_{\rm grav}^{(\tau)} \Psi_o(v) \nonumber \\
 &=:  \big(\Theta_0 + \Theta_1\big)\Psi_o(v).
 \label{eq:H-geo}
\end{align}
Here, $\Theta_0$ is the operator that appears in the $k=0$ quantum constraint in place of $\Theta$, while  $\Theta_1$ represents the additional term arising due to the positive spatial curvature in the $k=1$ LQC model. These operators are explicitly given by (cf. Appendix~\ref{app:DifferenceOp})
\begin{align}
\Theta_0\Psi_o(v) &= 
\frac{3\pi G}{4}
 \Big[(v+2)\sqrt{v(v+4)}\Psi_o(v+4) -2v \Psi_o(v)
 \nonumber \\ 
& \qquad  \qquad  \qquad +(v-2)  \sqrt{v(v-4)}\Psi_o(v-4)\Big],
\label{eq:Theta0b}
\\
\Theta_1\Psi_o(v) &=  \frac{3\pi G}{2} \left[\left(\sin^2 \left( \frac{\bar{\mu}\ell_o}{2} \right) - \frac{\bar{\mu}^2\ell_o^2}{4}\right)v^2 - \frac{\ell_o^2}{9}\left(\frac{v}{K}\right)^{4/3} \right] \Psi_o(v).
\label{eq:Theta1b}
\end{align}
Together, these terms determine the quantum evolution of the gravitational sector in LQC, providing a comprehensive description of the background dynamics.

As in the case of the spatially flat  FLRW model, the kinematical Hilbert space of the geometry is given by
\[
\mathscr{H}^o_{\rm kin}=L^2(\bar{\mathbb{R}}, d\mu_{\rm Bohr})\otimes L^2(\mathbb{R}, dT),
\]
where  $L^2(\bar{\mathbb{R}}, d\mu_{\rm Bohr})$ represents the gravitational Hilbert space  $\mathscr{H}_{\rm kin}^{\rm grav}$, and $L^2(\mathbb{R}, dT)$ corresponds to the Hilbert space $\mathscr{H}_T$ of the scalar field. Here,  $\bar{\mathbb{R}}$  denotes the Bohr compactification of the real line, and $d\mu_{\rm Bohr}$ is the Haar measure associated with it, reflecting the discrete nature of quantum geometry in LQC.
The physical states of the quantum geometry are those
$\Psi_o(v, T)\in \mathscr{H}^o_{\rm kin}$ lying in the kernel of $\hat{H}_{\rm geo}^{(\tau)}$. In other words, they are (the positive-frequency) solutions to the equation
\begin{equation}
-i\hbar \partial_T \Psi_o(v,T) = \hbar\sqrt{\Theta}\Psi_o(v,T) =: \hat{H}_o\Psi_o(v,T).
\label{eq:geometry-evol}
\end{equation}
These solutions determine our physical Hilbert space of the geometry $\mathscr{H}_{\rm phys}^o$ endowed with scalar product
\begin{equation}
\langle \Psi_o | \Psi_o^\prime\rangle_\varepsilon = \sum_{v\in \mathscr{L}_{\varepsilon}} \overline{\Psi_o(v, T_0)}\,  \Psi_o^\prime(v, T_0),
\end{equation}
at some instant $T_0$. Here,  $\mathscr{L}_{\varepsilon}=\mathscr{L}_{|\varepsilon|}\cup \mathscr{L}_{-|\varepsilon|}$, where $\mathscr{L}_{\pm|\varepsilon|}$ are lattices of points $\{\pm|\varepsilon|+4k; \, k\in\mathbb{Z}\}$  on the $v$ axis  \cite{Ashtekar:2006es}.

\section{Fermion modes on LQC background}
\label{sec:Fermion-LQC}

In this section, we aim to study the evolution of a Dirac fermion, treated as a perturbation, on the  FLRW background (\ref{eq:metric}). The Hamiltonian for the fermion field, as detailed in Sec.~\ref{sec:FermFLRW}, is expressed as an infinite set of Fermi oscillators, each associated with a harmonic mode $(n,p)$. To accomplish this, we analyze the full Hamiltonian constraint of the system, which includes both the gravitational and massless scalar-field sectors (representing the background geometry) coupled with the fermionic modes (representing the perturbations).

\subsection{Evolutionary equation for fermion modes}

As discussed in the previous section, to study the evolution of inhomogeneous fermionic modes $\{m_{np}$, $r_{np}$, $s_{np}$, $t_{np}\}$ on a loop quantum universe, 
we adopt the test-field approximation, in which the backreaction of these modes on the background quantum geometry is neglected. Physically, this is justified when the energy content of the fermionic perturbations remains small compared to the energy density of the homogeneous geometry. Mathematically, this allows a consistent ``truncation'' of the full theory, wherein only the zero mode of the scalar constraint is retained from the infinite hierarchy of quantum gravitational constraints. To implement this truncation, we smear the scalar constraint with a harmonic lapse function, effectively selecting a relational time variable---a homogeneous massless scalar field---to serve as a physical clock. The total Hamiltonian of the system in this gauge takes the form
\begin{equation}
H_{\rm tot} = H_{\rm geo}^{(\tau)} + H_{\rm ferm}^{(\tau)}.
\label{eq:constraint-ferm}
\end{equation} 
After quantizing the background geometry, we investigate the evolution of the quantum fermion field propagating on this quantized spacetime. By tracing out the gravitational degrees of freedom, we derive an evolution equation for the wave function of the fermion  modes,  $\psi_{np}$.

For the fermion field, the action of the total quantum Hamiltonian constraint $\hat{H}_{{\rm tot}}$ on a given state  $\Psi(T, v; x,y)$ in the full kinematical Hilbert space  $\mathscr{H}_{{\rm kin}}$ of the geometry-perturbation system, is expressed as
\begin{align}
\hat{H}_{{\rm tot}}\Psi &= \Big[\hat{H}_{\rm geo}^{(\tau)}\otimes \mathbb{I} + \sum_{np}\hat{H}_{np}^{(\tau)}(x, y)\Big]\Psi,
\label{eq:constraint}
\end{align}
where the quantized Hamiltonian $\hat{H}_{np}^{(\tau)}$ denotes the contribution from the quantized fermionic modes labeled by $(n,p)$ and includes both sets of mode functions $\{m_{np},s_{np}\}$ and $\{r_{np},t_{np}\}$, on the quantized FLRW background,
\begin{align}
\hat{H}_{np}^{(\tau)}& = \left[\lambda_n \ell^{-2}\hat{V}^{2/3}\otimes (\widehat{x\bar{x}} + \widehat{y\bar{y}})+m\ell^{-3}\hat{V}\otimes (\widehat{yx}+ \widehat{\bar{x}\bar{y}})\right].
\label{eq:Hamiltonian-SF-FLRW}
\end{align}
Using the explicit form of  $\hat{H}_{\rm geo}^{(\tau)}$ from Eq.~\eqref{eq:H-geo}, we rewrite the total constraint  \eqref{eq:constraint} as
\begin{align}
\hat{H}_{{\rm tot}}\Psi  &= \Big[-\frac{\hbar^2}{2\ell^3}\left(\partial_T^2+\Theta\right)\otimes \mathbb{I}  + \sum_{np}\hat{H}_{np}^{(\tau)}(x, y)\Big]\Psi.
\end{align}

In the test-field regime, this structure motivates a factorized ansatz for the total wave function $\Psi$  of the system,
\begin{align}
\Psi &:= \sum_{np}\Psi_{np}(T, v, m_{np}, s_{np}, r_{np}, t_{np})  \nonumber \\
& = \Psi_o(T, v)\otimes \prod_{np} \psi_{np}(T, m_{np}, s_{np}) \otimes  \psi_{np}(T, r_{np}, t_{np}) \nonumber \\
&=: \Psi_o(T, v)\otimes \prod_{np} \psi_{np}(T, x, y),
\label{eq:decompositionWF}
\end{align}
where $\Psi_o$ describes the background quantum geometry and each $\psi_{np}$ evolves independently.
Demanding that the total state $\Psi(T,v; x,y)$ be annihilated by the constraint  $\hat{H}_{{\rm tot}}\Psi\approx0$, we find that each mode's wave function, $\Psi_{np}=\Psi_o\otimes\psi_{np}$, must satisfy
\begin{align}
-\hbar^2\partial_T^2\Psi_{np} = \Big[\hat{H}^2_o - 2\ell^3\hat{H}_{np}^{(\tau)}\Big]\Psi_{np},
\label{eq:H-geo2}
\end{align}
where $\hat{H}_o^2=\hbar^2\Theta$ is proportional to the difference operator $\Theta$ that appears in the gravitational sector.
To define the physical inner product and ensure unitary evolution, we restrict attention to the positive-frequency sector of the theory. Assuming a suitable self-adjoint extension exists for the symmetric operator on the right-hand side of Eq.~\eqref{eq:H-geo2}, we identify it with $\hat{P}_T^2$, the square of the generator of evolution in relational time $T$. This yields the first-order Schr\"odinger-like evolution,
\begin{align}
-i\hbar\partial_{T}\Psi_{np} = \left[\hat{H}_o^2 - 2\ell^3\hat{H}_{np}^{(\tau)}\right]^{\frac{1}{2}}\Psi_{np}.
\label{eq:const-tot1}
\end{align}
This construction defines the truncated theory: a quantum FLRW universe coupled to test fermionic modes, where only the homogeneous constraint governs dynamics. Each fermionic mode evolves on the quantum-corrected background as if in an external field, and the total state of the system factorizes accordingly.

In this setting, the perturbative term {\small{}$2\ell^3\hat{H}_{np}^{(\tau)}$} may be treated as a small correction. To leading order, Eq.~\eqref{eq:const-tot1} simplifies via a series expansion of the square root,
\begin{align}
-i\hbar\partial_{T}\Psi_{np} &\approx \left[\hat{H}_o - \ell^3\hat{H}_o^{-\frac{1}{2}}\hat{H}_{np}^{(\tau)}\hat{H}_o^{-\frac{1}{2}}\right]\Psi_{np} \nonumber \\
&=:  \big(\hat{H}_o - \hat{H}_{np}^{(T)}\big)\Psi_{np},
\label{eq:const-tot1b}
\end{align}
which clearly separates the free gravitational evolution from fermionic mode corrections. Physical states $\Psi_{np}$  in the truncated theory belong to a Hilbert space $\mathscr{H}_{\rm phys}\equiv\mathscr{H}_{\rm kin}^o\otimes \mathscr{H}_{np}$  and are required to be normalizable under the physical inner product
\begin{align}
\langle\Psi_{np}, \Psi_{np}^\prime\rangle_\varepsilon &= \sum_{\varepsilon\in \mathscr{L}_\varepsilon}\int dx\, d\bar{x}\, dy\, d\bar{y}\,  e^{-x\bar{x} - y\bar{y}} \nonumber \\
&\qquad \qquad\times \overline{\Psi_{np}(T_0,v;x, y)}\, \Psi_{np}^\prime(T_0,v;x,y),
\end{align}
evaluated at any fixed value of the internal clock $T_0$.

\subsection{Interaction picture: Emerging dressed spacetimes} \label{sec:DressedBack}

To isolate the quantum evolution of the fermion modes on a ``time-dependent'' background, it is convenient to employ the interaction picture. This is achieved by introducing
\begin{equation}
\Psi_{{\rm int}}^{(np)}(v,x,y,T)=e^{-(i\hat{H}_{o}/\hbar)(T-T_{0})}\Psi_{np}(v, x,y,T),
\label{interaction-pic}
\end{equation}
where $T_0$ is an instant of the internal time. This implies that the geometry evolves according to $\hat{H}_o$, as given by Eq.~(\ref{eq:geometry-evol}), for any $\Psi_o \in \mathscr{H}_{{\rm kin}}^o$  in the Heisenberg picture,
\begin{equation}
\Psi_{o}(v,T)=e^{(i\hat{H}_{o}/\hbar)(T-T_{0})}\Psi_{o}(v,T_{0}).
\end{equation}
Plugging this into Eq.~(\ref{interaction-pic}), we obtain
\begin{equation}
\Psi_{{\rm int}}^{(np)}(v,x,y,T)=\Psi_{o}(v,T_{0})\otimes\psi_{np}(x,y,T),
\label{interaction-pic2}
\end{equation}
which indicates that the geometrical state in $\Psi_{{\rm int}}^{(np)}$ becomes frozen at the instant $T_0$. Consequently,  $\Psi_{{\rm int}}^{(np)}$ represents the evolution of the test spinor modes  $\psi_{np}(T, x, y)$ alone.

To decouple the evolution equations for the test spinor modes $\{m_{np}, s_{np}, r_{np}, t_{np}\}$, we substitute the wave function from Eq.~(\ref{interaction-pic2}) into the evolution equation (\ref{eq:const-tot1b}) and trace out the geometric state $\Psi_o(v, T_0)$. This procedure yields the quantum evolution equations for the light degrees of freedom $\psi_{np}$,
\begin{align}
i\hbar\partial_{T}\psi_{np}&= \langle\hat{H}_{np}^{(T)}\rangle_o\, \psi_{np}   \nonumber \\
&= \left[\lambda_n \ell^{-2}\langle \hat{H}_{o}^{-\frac{1}{2}}  \hat{V}^{\frac{2}{3}}\hat{H}_{o}^{-\frac{1}{2}} \rangle_o(\widehat{x\bar{x}} + \widehat{y\bar{y}}) + m\ell^{-3}\langle \hat{H}_{o}^{-\frac{1}{2}}  \hat{V}\hat{H}_{o}^{-\frac{1}{2}}  \rangle_o (\widehat{yx}+ \widehat{\bar{x}\bar{y}})\right]\psi_{np}\nonumber \\
&=:  \left[\lambda_n \langle \hat{\alpha} \rangle_o (\widehat{x\bar{x}} + \widehat{y\bar{y}}) + m\langle  \hat{\beta} \rangle_o (\widehat{yx}+ \widehat{\bar{x}\bar{y}})\right]\psi_{np},
\label{eq:Schrodinger-Q}
\end{align}
where we  define
\begin{equation}
\hat{\alpha}(T) := \ell^{-2}\hat{H}_o^{-\frac{1}{2}} \hat{V}^{2/3}(T) \hat{H}_o^{-\frac{1}{2}}  \quad \text{and} \quad 
\hat{\beta}(T)  := \ell^{-3}\hat{H}_o^{-\frac{1}{2}} \hat{V}(T) \hat{H}_o^{-\frac{1}{2}}.
\label{eq:alpha-beta}
\end{equation}
Here,   the expectation values $\langle\cdot\rangle_o$ are taken with respect to $\Psi_o(v, T_0)$. 
Note that, $\psi_{np}(T; x,y)$  in Eq.~(\ref{eq:Schrodinger-Q}) denotes the same wave function for both harmonic mode functions $\{m_{np}, s_{np}\}$ and $\{r_{np}, t_{np}\}$ in the product (\ref{eq:decompositionWF}). This formulation corresponds to the Heisenberg picture for quantum geometric elements, where the geometric state $\Psi_o(v, T_0)$   is frozen at time $T_0$, while the geometric operator, encoded in $\hat{V}(T)$, evolves with $T$,
\begin{align}
\hat{V}(T) &=e^{-(i\hat{H}_{o}/\hbar)(T-T_{0})}\, \hat{V}(T_0)\, e^{(i\hat{H}_{o}/\hbar)(T-T_{0})}. 
\end{align}
Thus, analogous to the classical theory [cf. Eq.~\eqref{eq:Schrodinger-Class0}], Eq.~(\ref{eq:Schrodinger-Q}) describes the evolution of the spinor mode states  $\psi_{np}$ on a time-dependent (classical) background.

Suppose that the Schr\"odinger equation (\ref{eq:Schrodinger-Q}) governs the evolution of spinor field modes on a classical isotropic spacetime, described by the metric
\begin{align}
\bar{g}_{\mu\nu}dx^\mu dx^\nu = -\bar{N}_{T}^2(T)dT^2 + \bar{a}^2(T)d\Omega^2_3.
\label{eq:metric-dressed}
\end{align}
On this background, following the classical Hamiltonian (\ref{eq:Schrodinger-Class1-A}), the evolution of the spinor field modes is given by
\begin{align}
i\hbar\partial_{T}\psi_{np}&=  \bar{N}_T \left[\lambda_n\bar{a}^{-1}(\widehat{x\bar{x}} + \widehat{y\bar{y}})+m(\widehat{yx}+ \widehat{\bar{x}\bar{y}})\right]\psi_{np}.
\label{eq:evol-eq1-dressed}
\end{align}
Comparing Eqs.~(\ref{eq:evol-eq1-dressed}) and (\ref{eq:Schrodinger-Q}) leads to the system
\begin{subequations}
\label{eq:main-sys}
\begin{align}
\lambda_n\bar{N}_{T}\bar{a}^{-1} &= \lambda_n\ell^{-2}\langle \hat{H}_{o}^{-\frac{1}{2}}  \hat{V}^{2/3} \hat{H}_{o}^{-\frac{1}{2}} \rangle_o,
\label{eq:main-sys-1}\\
m\bar{N}_{T} &=  m\ell^{-3} \langle \hat{H}_{o}^{-\frac{1}{2}}  \hat{V} \hat{H}_{o}^{-\frac{1}{2}} \rangle_o.
\label{eq:main-sys-2} 
	\end{align}
\end{subequations}
For massive fermion modes, this system consists of two equations for two unknowns, $(\bar{N}_{T}, \bar{a})$.  Solving for these variables, we obtain
\begin{subequations}
\label{eq:main-sys-solGen}
\begin{align}
\bar{N}_{T}(T) &=  \ell^{-3}\langle \hat{H}_{o}^{-\frac{1}{2}}  \hat{V}(T) \hat{H}_{o}^{-\frac{1}{2}} \rangle_o,
\label{eq:main-sys-sol1}
\\
\bar{a}(T) &= \ell^{-1}\frac{\langle \hat{H}_{o}^{-\frac{1}{2}}  \hat{V}(T) \hat{H}_{o}^{-\frac{1}{2}} \rangle_o}{\langle \hat{H}_{o}^{-\frac{1}{2}}  \hat{V}^{2/3}(T) \hat{H}_{o}^{-\frac{1}{2}} \rangle_o}\, .
\label{eq:main-sys-sol2}
\end{align}
\end{subequations}
These expressions provide a set of solutions for the components of the dressed metric (\ref{eq:metric-dressed}). 

For massless fermion modes, the system \eqref{eq:main-sys} reduces to a single equation governing the two metric functions 
 $(\bar{a}, \bar{N}_T)$. This reduction implies the existence of an infinite family of solutions, all constrained by the relation
\begin{equation}
\frac{\bar{N}_{T}}{\bar{a}} = \ell^{-2} \langle \hat{H}_{o}^{-\frac{1}{2}}  \hat{V}^{2/3}(T) \hat{H}_{o}^{-\frac{1}{2}} \rangle_o.
\label{eq:conformal-combination}
\end{equation}
The underdetermination of the system arises from the residual  freedom associated with conformal rescalings of the effective metric
$g_{\mu\nu} \to \Omega^2(x^\sigma) g_{\mu\nu}$, under which the massless Dirac equation (\ref{eq:DiracFull}) remains invariant. Consequently, massless fermions couple only to the conformally invariant combination $\bar{N}_T/\bar{a}$, rather  than  to $\bar{a}(T)$ or  $\bar{N}_T(T)$ individually.
To elucidate this behavior, we introduce a dressed conformal time,
\begin{align} 
d\bar{T} \equiv \frac{\bar{N}_{T}}{\bar{a}} dT = \ell^{-2} \langle \hat{H}_{o}^{-\frac{1}{2}}  \hat{V}^{2/3}(T) \hat{H}_{o}^{-\frac{1}{2}} \rangle_o\, dT,
\label{eq:time-rescaling}
\end{align} 
which allows us to express the effective line element in the form
\begin{align} 
\bar{g}_{\mu\nu} dx^\mu dx^\nu &= \bar{a}^2(T) \left[ - d\bar{T}^2 + d\Omega^2_3 \right].
\end{align} 
Thus, in the ultrarelativistic regime, the physically relevant geometry reduces to the conformal equivalence class 
\begin{align} 
\bar{g}_{\mu\nu}^\prime dx^\mu dx^\nu = - d\bar{T}^2 + d\Omega^2_3.
\end{align} 
Quantum gravitational effects manifest exclusively through the temporal reparametrization $T\to \bar{T}$, which encodes expectation values of geometric operators. While the spatial three-sphere geometry remains classical, the temporal component undergoes a quantum-corrected rescaling, resulting in a hybrid spacetime where quantum effects modify only the temporal evolution.
This demonstrates that, for massless modes, only the ``causal structure''---not the full metric---carries physical significance. Moreover, the conformal flatness of the background precludes particle-antiparticle production for massless fermions, as such processes would require symmetry-breaking terms (e.g., fermion masses or nonconformal interactions) to generate detectable pair creation.

\section{Backreaction effects}
\label{sec:BReffects}

In the previous section, we analyzed the evolution of fermion modes within the test-field approximation, neglecting their backreaction on the quantum geometry. However, to develop a more complete and self-consistent framework, it is essential to incorporate the effects of this backreaction \cite{Parvizi:2021ekr,Graef:2020qwe,SVicente:2022ebm}. In what follows, we investigate the dynamics of fermion modes on a quantum gravitational background that has been modified to include the influence of fermionic backreaction.

The  backreaction effects arise from the energy expectation values of fermionic eigenstates {\small{}$\psi_{np}^{(I)}$}, which include (i) a negative-energy vacuum state ({\small{}$E_n^{(0)}=-w_n; \psi^{(0)}_{np}$}), (ii) two degenerate zero-energy modes ({\small{}$\psi^{(1,2)}_{np}$}), and (iii) a positive-energy excited state ({\small{}$E_n^{(3)}=+w_n; \psi^{(3)}_{np}$}). While this energy spectrum differs from the conventional Dirac structure in $(3+1)$-dimensional spacetime, it retains essential physical characteristics. Additionally, transitions between the vacuum and excited states of massive modes ($\psi^{(0)}_{np} \rightarrow \psi^{(3)}_{np}$) represent particle-antiparticle creation and annihilation processes that can also contribute to the dynamical evolution of the quantum background. The zero-energy states, by contrast, are nondynamical artifacts of the finite-dimensional truncation inherent in the Grassmann-algebraic description and are expected to decouple in a complete quantum field-theoretic treatment.

\subsection{Born-Oppenheimer approximation}

A well-established method for incorporating  backreaction is the Born-Oppenheimer approximation, which separates the dynamics of the light (fermionic) and heavy (geometric) degrees of freedom \cite{Dapor:2012jg, Parvizi:2021ekr}. Under this approximation, the total wave function of the system (for each fermionic mode) is factorized as a tensor product,
\begin{align}
\tilde{\Psi}_{np}(T, v; x, y)
& = \tilde{\Psi}_{np}^o(T, v)\otimes \psi_{np}(T, x, y),
\label{eq:decompositionWF-BR}
\end{align}
where $\tilde{\Psi}_{np}^o(T, v)$ describes the quantum geometry modified by the fermion-induced backreaction (\ref{eq:omega-energy}) in the operator sense, and $\psi_{np}(T, x, y)$ encodes the fermionic degrees of freedom.

The fermionic and geometric sectors, now, are governed by two coupled equations. The fermionic eigenstates $\psi_{np}$ and their associated eigenvalues $E_{n}$ are obtained from the stationary equation  (\ref{eq:Eigenvalue1}),  while the evolution of the quantum geometry is governed by the effective Hamiltonian constraint,
\begin{align}
\hat{H}_{{\rm tot}}\tilde{\Psi}_{np}^o &= \Big[\hat{H}_{\rm geo}^{(\tau)} +  \langle\hat{H}_{np}^{(T)}(v)\rangle_\psi\Big]\tilde{\Psi}_{np}^o \nonumber \\
&= \Big[-\frac{\hbar^2}{2\ell^3}(\partial_T^2+\Theta) +  \langle\hat{H}_{np}^{(T)}(v)\rangle_\psi\Big]\tilde{\Psi}_{np}^o.
\label{eq:constraintBRTerm-A}
\end{align}
In this equation, we adopt a fully quantum geometric description, when including the fermionic contribution, by promoting the geometric variables  to operators, i.e., $V \to \hat{V}$. 
As a result, $E_n(v)$ becomes the operator-valued quantity $\hat{E}_n(v)$, expressed as
\begin{equation}
\hat{E}_{n}(v) = \langle \hat{H}_{np}^{(T)}(v)\rangle_\psi = \pm\sqrt{\frac{\lambda_n^2}{\ell^4} \hat{V}^\frac{4}{3} + \frac{m^2}{\ell^6}\hat{V}^2},
\label{eq:operatorV-E}
\end{equation}
where the minus and plus signs correspond to the vacuum and excited (pair) state energy contributions, respectively. 
This operator acts diagonally on the volume eigenbasis.
Furthermore, since $\hat{E}_{n}$  depends only on the  mode label $n$ (and not $p$), the modified background wave function depends solely on $n$. So, we will henceforth drop  the label $p$ in $\tilde{\Psi}_{np}^o$.

Solving this quantum constraint yields the evolution equation,
\begin{align}
-\partial_T^2\tilde{\Psi}_{n}^o &=\Big[\Theta - \frac{2\ell^3}{\hbar^2}\hat{E}_n\Big]\tilde{\Psi}_{n}^o  \nonumber\\
&=: (\Theta+\Theta_n) \tilde{\Psi}_{n}^o =  \tilde{\Theta}_n\tilde{\Psi}_{n}^o,
\label{eq:constraintBRTerm5-A}
\end{align}
where we define the mode-dependent,  Hamiltonian operator of backreaction as
\begin{equation}
\Theta_n=-\frac{2\ell^3}{\hbar^2}\hat{E}_n,
\label{eq:Theta-n}
\end{equation}
in analogy with the  gravitational Hamiltonian $\Theta=-(2\ell^3/\hbar^2)\hat{H}_{\rm grav}^{(\tau)}$ [cf. Eq.~(\ref{eq:H-geo})].

The corresponding positive-frequency solution is given by
\begin{align}
-i\hbar\partial_T\tilde{\Psi}_n^o = \Big[\hat{H}_o^2 - 2\ell^3\hat{E}_n\Big]^\frac{1}{2}\tilde{\Psi}_n^o,
\label{eq:constraintBRTerm6-A}
\end{align}
which can be approximated as
\begin{align}
-i\hbar\partial_T\tilde{\Psi}_{n}^o(T, v) &\approx \Big[\hat{H}_o - \hat{E}_n^{(T)}\Big]\tilde{\Psi}_{n}^o(T, v).
\label{eq:BO2}
\end{align}
Here, the backreaction operator  $\hat{E}_n^{(T)}$,
\begin{align}
\hat{E}_n^{(T)}(v) &:= \ell^3\hat{H}_o^{-\frac{1}{2}}\hat{E}_n(v)\hat{H}_o^{-\frac{1}{2}} \nonumber \\
&\, = \pm \left(\lambda_n^2\ell^{2}\hat{H}_o^{-1} \hat{V}^\frac{4}{3}\hat{H}_o^{-1} +m^2\hat{H}_o^{-1}\hat{V}^2\hat{H}_o^{-1}\right)^{1/2},
\label{eq:EnergyOperator}
\end{align}
encapsulates the leading‐order fermionic correction. This separation is valid provided the characteristic fermion scales  {\small{}$\lambda_n\ell\hat{H}_o^{-1/2} \hat{V}^\frac{2}{3}\hat{H}_o^{-1/2}$} and {\small{}$m\hat{H}_o^{-1/2} \hat{V}\hat{H}_o^{-1/2}$} remain small compared to $\hat{H}_o$ ensuring geometry (the heavy sector) and fermions (the light sector) decoupled.

Equation~(\ref{eq:Theta-n}) illustrates   that  the sign  of the backreaction operator $\Theta_n$ hinges on which fermionic eigenstates are occupied. 
In the vacuum state, with energy $-|E_n|$, the fermion mode $n$ 
contributes positively to the effective Hamiltonian (\ref{eq:constraintBRTerm5-A}), yielding  $\Theta_n=(2\ell^3/\hbar^2)|\hat{E}_n|>0$.
In contrast, populating the pair state (with energy $+|E_n|$)  flips the sign, resulting in a negative contribution to the Hamiltonian: $\Theta_n=-(2\ell^3/\hbar^2)|\hat{E}_n|<0$.
A transition of a single mode from the vacuum state to the excited (pair) state changes  $E_n$ by $+2w_n$, inducing a jump in  $\Theta_n$  of  $-(4\ell^3/\hbar^2)|\hat{E}_n|$.  
Conversely, returning the mode to the vacuum state adds  $+(4\ell^3/\hbar^2)|\hat{E}_n|$ back to the Hamiltonian.
In what follows, we will analyze the spectrum of the backreacted geometry and discuss its physical implications for the early Universe.

\subsection{Spectrum of the backreacted geometry}

In the absence of backreaction effects, the evolution of the quantum geometry is governed by Eq.~\eqref{eq:geometry-evol}. In this setting, the difference operator $\Theta$ is self-adjoint on a separable, countable‐basis Hilbert space  $\mathscr{H}^{\rm grav}_{|\varepsilon|}$. Moreover,  its spectrum is discrete and nondegenerate,  with eigenfunctions $e_k(v)$ satisfying
\begin{equation} 
\Theta\, e_k(v) = \omega_k^2\, e_k(v), \quad \text{with} \quad \langle e_k | e_{k'} \rangle = \delta_{k,k'}. 
\label{eq:eigenvalue-unperturbed}
\end{equation} 
These eigenfunctions form a complete orthonormal basis for  $\mathscr{H}^{\rm grav}_{|\varepsilon|}$, and the general solution to the quantum constraint \eqref{eq:geometry-evol} is given by
\begin{equation} 
\Psi_o(v, T) = \sum_k c_k^{o}\, e_k(v)\, e^{i\omega_k T},
\label{eq:FullunperturbedWF}
\end{equation} 
where the coefficients $c_k^o$ are square summable, ensuring normalizability of $\Psi_o$. Physical states are constructed to respect orientation-reversal symmetry under $v \to -v$, and the most general solution includes both positive- and negative-frequency components, in line with the full solution space of the quantum constraint.
The  eigenbasis $e_{k}(v)$ captures essential features of LQC in a closed universe, including the resolution of singularities via a quantum bounce and semiclassical behavior at large volumes  \cite{Ashtekar:2006es}.

When a fermion mode labeled by $(n, p)$ backreacts on the background quantum geometry, it contributes a perturbative energy  $\Theta_n$ [cf. Eq.~(\ref{eq:Theta-n})] with a sign depending on which state it occupied.  This effect modifies the gravitational sector by introducing an additional, mode-dependent potential, deforming the original eigenfunctions $e_k(v)$ into a new set $e_k^{(n)}(v)$. The perturbed quantum state $\tilde{\Psi}_n^{o}(v, T)$ then satisfies the modified eigenvalue equation
\begin{equation}
\big[\Theta + \Theta_n\big]e_{k}^{(n)} (v) = (\omega_{k}^{(n)})^2\, e_{k}^{(n)} (v), \quad \text{with} \quad \langle e_{k}^{(n)}| e_{k'}^{(n)}\rangle=\delta_{k, k'}.
\label{eq:EigenvalueGeo-pert}
\end{equation}
Here, $\Theta_n(v)$ acts diagonally in the volume basis as an effective potential, encoding exactly how the fermion's energy content reshapes the underlying quantum geometry.
Because  $\Theta$ and $\Theta_n$  are diagonal in the same basis, they commute and their sum remains self-adjoint, guaranteeing real eigenvalues. Moreover, since $|\Theta_n|$ ($\sim\hat{V}$) grows no faster than $\Theta$ ($\sim \hat{V}^{4/3}$) at large volume $v$---making it a relatively bounded (in fact, relatively compact) perturbation---the combined operator $(\Theta+\Theta_n)$ preserves the compactness of the resolvent and thus retains a discrete spectrum. Analytic perturbation theory, under the usual ``gap conditions,'' then ensures these eigenvalues remain nondegenerate---any level shifts continuously and avoids crossings unless a symmetry compels a degeneracy \cite{Kato:1966}. In practice, each $\omega_k^{(k)}(v)$  may shift and each  $e_k^{(n)}(v)$ deform in a mode-dependent way, especially in the high-curvature regime near the bounce, but the overall spectral structure---self-adjointness, discreteness, and nondegeneracy---remains intact.

The eigenvalue equation (\ref{eq:EigenvalueGeo-pert}) behaves analogous to a stationary Schr\"odinger equation describing a particle in a finite potential well, given by  $U_{\rm eff}(v)=U_o(v)+\Theta_n(v)$, where $U_o(v)$ represents a positive potential arising from the unperturbed background quantum geometry, $\Theta\sim -\partial_v^2 + U_o(v)$,  within a semiclassical Wheeler-DeWitt (WDW) framework. Within the well---around the quantum bounce, where the potential peaks at $U_{\rm eff}(v_{\rm b})=\omega^2_k$---the eigenfunctions  $e_k(v)$ exhibit oscillatory behavior [$U_{\rm eff}(v)<\omega_k^2$], while outside [$U_{\rm eff}(v)>\omega_k^2$] they decay. The fermionic backreaction term $\Theta_n$ then shifts this minisuperspace barrier up or down, depending on mode occupation.
In the vacuum state,  we have  $\Theta_n>0$,  which introduces an additional positive contribution to the effective potential. This raises the barrier peak and shifts it to a smaller volume, $v_{\rm b}$.
Equivalently, within the effective Friedmann picture,  the bounce condition  $\rho_T-|\rho_n|=\rho_{\rm cr}$ is satisfied at a higher background energy density $\rho_{T}$. This implies that the quantum bounce is delayed---occurring later in the contraction phase of the Universe. The barrier also broadens, making the classically forbidden region less abrupt \cite{Motaharfar:2022pjp}.
By contrast, when the pair state is occupied, $\Theta_n<0$, which leads to  a lower effective potential: $U_{\rm eff}=U_o(v)-|\Theta_n(v)|<U_{o}(v)$. This reduces the height of the barrier and shifts its peak to a larger volume $v_{\rm b}$. Consequently, the bounce occurs at a lower background energy density $\rho_T$, since the condition  $\rho_T+\rho_n=\rho_{\rm cr}$ is now satisfied earlier during the contraction phase---and the bounce itself becomes more abrupt.

A single-mode transition from a vacuum state to a particle-antiparticle state requires an energy input of $+2|E_{n}|$, which  induces a negative jump $\Delta\Theta_n=-(4\ell^3/\hbar^2)|\hat{E}_{n}|$. This lowers the effective minisuperspace potential barrier and sharply advances the quantum bounce.
Conversely, when the mode relaxes back to its vacuum state, the shift reverses to $\Delta\Theta_n=+(4\ell^3/\hbar^2)|\hat{E}_{n}|$, which raises and broadens the barrier, restoring the delayed and smoother bounce characteristic of the vacuum-dominated dynamics.
Physically, the energy for these nonadiabatic transitions is supplied by the rapid, time-dependent evolution of the quantum background, whose changing scale factor ``pumps'' fermionic modes out of their instantaneous vacuum, in direct analogy to Parker particle production in curved spacetime \cite{Parker:1968mv}.
Moreover, since the characteristic frequency scales as $E_n\sim n$ at large mode number, high-frequency modes induce proportionally larger $|\Delta\Theta_n|$ shifts than low-frequency ones. This linear scaling enables a fine-grained rainbow of bounces---each fermion mode $n$ selects its own bounce volume, timing, and sharpness when excited.

In the deep Planckian regime ($V\to V_{\rm b}$), the  effective potential reduces to 
\begin{equation}
\Theta_n(v) \sim \pm\lambda_n \ell^{-2} \hat{V}^{2/3}.
\label{eq:1stapprox}
\end{equation}
This  introduces an  $n$-dependent potential well  that yields  oscillatory eigenfunctions around the bounce and exponential decay elsewhere,  giving rise to nonuniform level spacing and interference patterns that encode the fermionic sector's structure.  At intermediate volumes, high-$n$ modes can still amplify this term to rival or exceed the mass contribution, so ultraviolet fermions induce appreciable corrections even outside the Planck regime. 
By contrast, in the large-volume semiclassical limit and for low-lying modes, the potential is dominated by
\begin{equation}
\Theta_n(v) \sim  \pm m \ell^{-3} \hat{V},
\label{eq:2ndapprox}
\end{equation}
suppressing quantum gravity corrections and recovering classical WDW behavior.

The current treatment is limited to a qualitative exploration aimed at highlighting the principal effects of fermionic backreaction. A comprehensive numerical analysis of these spectral modifications lies beyond the scope of this work. A more detailed, quantitative investigation will be presented in a forthcoming study.

\subsection{Rainbow dressed background}

To systematically characterize the complete wave function $\tilde{\Psi}_n^o(v, T)$ and its associated eigenbasis {\small{}$e^{(n)}_{k}(v)$}, we adopt a perturbative approach that explicitly captures the hierarchy of corrections. The perturbed eigenfunctions can be expanded as
\begin{equation}
|e^{(n)}_{k}\rangle = \underline{N}_k\left[|e_{k}\rangle + |\delta e^{(n)}_{k}\rangle\right],
\label{eq:xi-mu}
\end{equation}
where \(\underline{N}_k\) is a normalization factor ensuring the orthonormality of the resulting eigenbasis.  A first-order approximation to the solution of  (\ref{eq:EigenvalueGeo-pert}) is then constructed using a fixed spectral profile  \(c_k^{(n)}\) and the corrected eigenfunctions  {\small{}$e^{(n)}_{k}(v)$}, yielding
\begin{equation}
\tilde{\Psi}_{n}^o(v, T) = \sum_{k} c_k^{(n)} \, e^{(n)}_{k}(v) \, e^{i\omega_{k}T},
\label{eq:tildePsi-BR}
\end{equation}
where the function {\small{}$e^{(n)}_{k}(v)$}  satisfies the perturbed eigenvalue equation \eqref{eq:EigenvalueGeo-pert}.  In the regime where fermion-geometry correlations remain negligible, substituting the perturbative expansion \eqref{eq:xi-mu} into Eq.~\eqref{eq:tildePsi-BR} gives
\begin{align}
\tilde{\Psi}_{n}^o(v, T) &= \sum_k c_k\, e_{k}(v)\, e^{i\omega_{k}T} + \sum_k c_k\, \delta e^{(n)}_{k}(v)\, e^{i\omega_{k}T} \nonumber \\
&=: \Psi_{o}(v, T) + \delta\Psi_{n}(v, T).
\label{eq:Psi1}
\end{align}
Here, we assume the same spectral coefficients $c_k=c_k^{(n)}=c_k^{o}$ for both the unperturbed and perturbed components. In this decomposition, the first term, $\Psi_o$,  represents the unperturbed quantum geometry, while the second term, $\delta\Psi_n$, encodes first-order corrections arising from the backreaction of the $n$th fermion mode.

Upon incorporating the first-order correction (\ref{eq:Psi1})  into the composite wave function (\ref{eq:decompositionWF-BR}), within the Born-Oppenheimer approximation, and substituting this expansion into the total constraint (\ref{eq:constraintBRTerm-A}), we derive the evolution equation for $\tilde{\Psi}_{np}=\tilde{\Psi}_n^o\otimes\psi_{np}$ as
\begin{equation}
-i\hbar\partial_T\tilde{\Psi}_{np} = \big(\hat{H}_o - \hat{H}_{np}^{(T)}\big)\tilde{\Psi}_{np}.
\end{equation}
By tracing out the geometric degrees of freedom using the modified  state $\tilde{\Psi}_n^o(v, T_0)$ (within the intercation picture), the dynamics of the fermionic modes are modified to
\begin{align}
i\hbar\partial_{T}\psi_{np}&=  \left[\lambda_n \ell^{-2}\langle \hat{H}_{o}^{-\frac{1}{2}} \hat{V}^{\frac{2}{3}}\hat{H}_{o}^{-\frac{1}{2}} \rangle(\widehat{x\bar{x}} + \widehat{y\bar{y}})+m\ell^{-3}\langle \hat{H}_{o}^{-\frac{1}{2}}  \hat{V}\hat{H}_{o}^{-\frac{1}{2}}  \rangle (\widehat{yx}+ \widehat{\bar{x}\bar{y}})\right]\psi_{np}.
\label{eq:Schrodinger-BR}
\end{align}
The expectation values now evaluated on the perturbed geometric state $\tilde{\Psi}_{n}^o = \Psi_{o} + \delta\Psi_{n}$. This leads to a natural decomposition of geometric operator expectation values,
\begin{align}
\langle \hat{O} \rangle &= \langle \hat{O} \rangle_o + \langle \hat{O} \rangle_n,
\label{eq:expectation_decomp}
\end{align}
where $\hat{O}$ stands for $\hat{\alpha}(T)$ or $\hat{\beta}(T)$, defined in Eq.~(\ref{eq:alpha-beta}). The correction term is given by
\begin{align}
\langle \hat{O}\rangle_n &:=  \langle\Psi_o| \hat{O} |\delta\Psi_{n}\rangle + \langle\delta\Psi_{n}| \hat{O}|\Psi_o\rangle   
+ \langle\delta\Psi_{n}| \hat{O} |\delta\Psi_{n}\rangle,
\end{align}
capturing the influence of backreaction on the dressed geometry. 
These  corrections are mode-dependent due to the explicit $n$ dependence of the perturbation wave function,
\begin{equation}
\delta\Psi_n(v, T)=\sum_k c_k\, \delta e_{k}^{(n)}(v)\, e^{i\omega_kT},
\end{equation}
which introduces $n$ dependence  into the geometric observables  $\langle\hat{\alpha}\rangle_n$ and $\langle\hat{\beta}\rangle_n$.
Consequently, the effective Schr\"odinger equation governing fermionic dynamics acquires the refined form
\begin{align}
i\hbar\partial_{T}\psi_{np}&=  \left[\lambda_n \big(\langle\hat{\alpha} \rangle_o + \langle\hat{\alpha} \rangle_n\big)(\widehat{x\bar{x}} + \widehat{y\bar{y}})+m\big(\langle\hat{\beta} \rangle_o + \langle\hat{\beta} \rangle_n\big) (\widehat{yx}+ \widehat{\bar{x}\bar{y}})\right]\psi_{np}.
\label{eq:Schrodinger-BR2}
\end{align}

The effective evolution equation~\eqref{eq:Schrodinger-BR2} governs the dynamics of each fermionic perturbation mode $(n,p)$ propagating on a dressed quantum-corrected background metric,
\begin{align}
\tilde{g}_{\mu\nu}dx^\mu dx^\nu = -\tilde{N}_{T}^2(T)dT^2 + \tilde{a}^2(T)d\Omega^2_3.
\label{eq:metric-dressedBR}
\end{align}
In fact, a direct comparison between the evolution equation (\ref{eq:Schrodinger-BR2}) and the  Schr\"odinger equation (\ref{eq:Schrodinger-Class0}) associated with the  metric  (\ref{eq:metric-dressedBR}) reveals how the metric components $(\tilde{N}_T, \tilde{a})$ are encoded in expectation values of quantum geometric operators. Specifically, we obtain
\begin{subequations}
\label{eq:main-sysBR}
\begin{align}
\lambda_n\tilde{N}_{T}\tilde{a}^{-1} &= \lambda_n\langle \hat{\alpha}(T) \rangle_o\left(1+\delta_{n}^{(1)}\right),
\label{eq:main-sys-1BR}\\
m\tilde{N}_{T} &=  m \langle \hat{\beta}(T) \rangle_o \left(1+\delta_{n}^{(2)}\right),
\label{eq:main-sys-2BR} 
	\end{align}
\end{subequations}
where the backreaction-induced correction terms  are quantified by
\begin{align}
\delta_{n}^{(1)}(T) \equiv \frac{\langle \hat{\alpha}(T) \rangle_n}{\langle \hat{\alpha}(T) \rangle_o} 
\quad \text{and} \quad 
\delta_{n}^{(2)}(T) \equiv \frac{\langle \hat{\beta}(T) \rangle_n}{\langle\hat{\beta}(T) \rangle_o}.
\end{align}
Crucially, these corrections carry explicit dependence on the fermionic mode $n$ due to the structure of the perturbative corrections $\delta\Psi_n$, thus imprinting spinor-mode dependence onto the background geometry itself.

For massive fermion modes, the system~\eqref{eq:main-sysBR} uniquely determines the components of the dressed metric. Solving for $\tilde{N}_T$ and $\tilde{a}$, we find
\begin{subequations}
\label{eq:main-sys-solBRGen}
\begin{align}
\tilde{N}_{T}(T) &= \bar{N}_{T}(T)F_{n}(T),
\label{eq:main-sys-sol1BR}
\\
\tilde{a}(T) &= \bar{a}(T)G_{n}(T),
\label{eq:main-sys-sol2BR}
\end{align}
\end{subequations}
where $(\bar{N}_T, \bar{a})$  are the unperturbed dressed metric components [defined in Eq.~\eqref{eq:main-sys-solGen}], and the mode-dependent  functions $F_n$ and $G_n$ encode backreaction effects via
\begin{subequations}
\label{eq:main-sys-solBRGen2}
\begin{align}
F_{n}&\equiv \left[\left(1+\delta_{n}^{(1)}\right)\left(1+\delta_{n}^{(2)}\right)^3\right]^{\frac{1}{4}}, \\
G_{n}&\equiv \left(\frac{1+\delta_{n}^{(2)}}{1+\delta_{n}^{(1)}}\right)^{\frac{1}{4}}.
\end{align}
\end{subequations}
In the test-field limit, where backreaction effects vanish   ({\small{}$\delta_{n}^{(1)}, \delta_{n}^{(2)}\to 0$}), the modulation functions tend toward unity  ($F_{n}, G_{n}\to 1$),  restoring the unperturbed geometry. In contrast,  Eqs.~\eqref{eq:main-sys-solBRGen} demonstrate that backreaction renders the effective geometry mode dependent: each fermion mode $n$ experiences a distinct spacetime. This mode dependence constitutes a rainbow metric structure---a hallmark of genuinely quantum gravitational effects beyond semiclassical treatments.

In the massless case, the system~\eqref{eq:main-sysBR} reduces to a single equation,
\begin{equation}
\frac{\tilde{N}_{T}}{\tilde{a}} = \langle \hat{\alpha}(T) \rangle_o\left(1+\delta_{n}^{(1)}\right).
\label{eq:main-sys-1BR-massless}
\end{equation}
Here, no unique solution of the full background metric emerges. Instead, the geometry is determined only up to a conformal factor that is mode dependent. The corresponding dressed metric takes the form 
\begin{align} 
\tilde{g}_{\mu\nu} dx^\mu dx^\nu &= \tilde{a}^2(T) \left[ - d\tilde{T}^2 + d\Omega^2_3 \right] \nonumber \\
&=: \tilde{a}^2(T)\, \tilde{g}_{\mu\nu}^\prime dx^\mu dx^\nu,
\end{align} 
where the conformal time coordinate $\tilde{T}$  is related to the original (relational) time $T$ or  the dressed time $\bar{T}$ [cf. Eq.~(\ref{eq:time-rescaling})] via
\begin{align} 
d\tilde{T} \equiv \frac{\tilde{N}_{T}}{\tilde{a}} dT = \left(1+\delta_{n}^{(1)}\right) d\bar{T}.
\label{eq:time-rescalingBR}
\end{align} 
Thus, in the massless limit, only the conformal class  $\tilde{g}'_{\mu\nu}(\tilde{T})$ of  metrics  has physical relevance. The scale factor $\tilde{a}(T)$  becomes irrelevant for massless fermions, which are sensitive only to the conformal structure of spacetime. As in the case without backreaction, the spatial geometry remains unchanged, but quantum gravitational corrections---now including backreaction---induce a nontrivial rescaling of time, dependent on the fermionic mode. Consequently, each massless mode evolves along a distinct conformal time, leading to a mode-dependent conformal geometry, $\tilde{g}_{\mu\nu}^\prime(\tilde{T})$.

\subsection{Comparison with bosonic backreaction}

The backreaction of bosonic and fermionic fields on a quantum-cosmological background shares the common feature of introducing mode-dependent modifications to the effective geometry, but the two cases differ in both structure and physical origin. For a scalar (bosonic) field on a flat FLRW background \cite{Parvizi:2021ekr}, each Fourier mode $\mathbf{k}$ gives rise to an infinite tower of energy eigenvalues
\begin{equation} 
\epsilon^{(s)}_{\mathbf{k}}(v) = \left(s + \tfrac{1}{2}\right) \hbar k \hat{V}^{-1/3}, 
\end{equation} 
so that every excitation level $s$ (with $s=0, 1, 2, \ldots$) contributes a distinct, negative perturbation, $\Theta_n<0$. As a result, each bosonic mode ``sees'' a slightly different effective metric, parametrized continuously by both momentum  $\mathbf{k}$ and excitation level $s$---the hallmark of a continuous rainbow geometry. The vacuum ($s=0$) alone carries positive, zero‐point energy $\frac{1}{2}\hbar kV^{1/3}$, which (similar to excited states) lowers the minisuperspace potential barrier and shifts the bounce to occur at lower density ($\rho_T=\rho_{\rm cr}-\rho_s$) or larger volume, effectively advancing the bounce in the contraction history.

By contrast, fermionic fields on a closed FLRW background quantize into a finite Grassmann algebra: each spinor mode $(n,p)$ has only a negative-energy vacuum state, two zero-energy states (artifacts of the Grassmann truncation), and a positive-energy pair state. Only the vacuum and pair states carry nontrivial backreaction. In the vacuum, the negative eigenvalue $-|E_n|$ yields a positive shift $\Theta_n>0$, raising the bounce barrier and causing the Universe to bounce at a smaller volume (later in contraction), while in the pair state $+|E_n|$ flips  $\Theta_n<0$, lowering the barrier and producing a faster bounce at larger volume. 
Since each mode has only these two eigenenergies, the resulting fermionic rainbow geometry is inherently discrete, with exactly two possible backreacted geometries per mode. High-frequency modes (large $n$) induce much larger shifts than low-frequency ones, enabling mode-by-mode tuning of the bounce's volume, timing, and sharpness.

Finally, whereas the scalar-field analysis in \cite{Parvizi:2021ekr} was performed on a flat FLRW background, our fermionic study takes place on a closed FLRW geometry. Although the essential rainbow-type backreaction features persist at large volumes (where curvature is negligible), a fully parallel comparison would require extending the bosonic analysis to the closed-universe setting. Such an extension would deepen our understanding of how statistics---bosonic vs. fermionic---uniquely imprint quantum fields onto the early Universe's geometry.

\section{Early-Universe dynamics}
\label{Sec:EarlyUniverse}

In this section, we analyze how fermionic backreaction and the resulting rainbow geometry affect early-Universe dynamics. We quantify backreaction by the mode-dependent energy densities
\begin{align}
\rho_n(T) = \langle : \widehat{V^{-1}(T)E_n^{(T)}} : \rangle,
\label{eq:FermionicEnergyDensity}
\end{align}
evaluated on physical hypersurfaces of constant relational time $T$, 
where {\small{}$E_n^{(T)}$} is the physical Hamiltonian of the $n$th fermion mode [cf. Eq.~(\ref{eq:EnergyOperator})]. These backreaction effects are most pronounced in two regimes:
(i) near the quantum bounce ($V\approx V_{\rm b}$) and (ii) during the preinflationary expansion at large volume.

Initially, at the bounce, all fermion modes reside in their vacuum state.  As the Universe expands, nonadiabatic transitions can excite these modes, leading to particle-antiparticle pair production.  These excitations raise $\rho_n$, altering the total energy density as $\rho_T+\rho_n$. In this limit, one finds approximately
\begin{align}
\rho_n \approx 2\lambda_n\ell \langle \hat{H}_o^{-1/2} \hat{V}^{-1/3}(T) \hat{H}_o^{-1/2} \rangle.
\label{eq:EnergyOperator-1}
\end{align}
Compared with a massless scalar's scaling $\rho_T \sim \langle \hat{V}^{-2} \rangle_o$,  this fermionic contribution scales more mildly  $\sim \langle \hat{V}^{-1/3} \rangle_o$. Physically, these excited modes can shift the bounce volume (in the contracting phase), alter its timing, and drive a more rapid postbounce expansion.

Fermionic particle production also breaks the exact time symmetry of the LQC bounce. In the absence of backreaction, contraction and expansion mirror each other; however, excited fermions introduce (i) asymmetric particle production across the bounce surface and (ii) effective anisotropic stresses due to their spinor structure.
Together, these effects yield different effective equations of state before and after the bounce, typically enhancing the postbounce growth and endowing the early Universe with a preferred arrow of time. Crucially, this asymmetry requires a nonzero fermion mass: massless fermions retain conformal invariance in the background and therefore do not undergo gravitational particle production in this framework.

In the regime far from the bounce ($V \gg V_{\rm b}$), the Universe's geometry becomes semiclassical. In this limit, the expectation value $\langle \hat{H}_o^{-1} \rangle_o$ stabilizes along the classical trajectory. Specifically, in the unperturbed geometry, physical states satisfy $\hat{H}_o \Psi_o = P_T^2 \Psi_o$, where $P_T$ is a constant of motion. For states sharply peaked around a classical trajectory, this leads to
\begin{equation}
\langle \hat{H}_o^{-1}\rangle_o = P_T^{-2}=\text{const}.
\end{equation}
Consequently, the backreaction energy density from a massive fermion mode becomes
\begin{align}
\rho_n &= m\langle\hat{H}_o^{-1}\rangle_o \approx \text{const.}
\label{eq:EnergyOperator-2}
\end{align}
This constant energy density effectively acts as a cosmological constant, potentially driving late-time accelerated expansion without the need for an explicit scalar potential.   In contrast, massless fermions yield an energy density that diminishes with the Universe's expansion, rendering them incapable of sourcing such acceleration.

In the unperturbed background,  the scalar‐field momentum $P_T$  and its energy density $\rho_T$ are related by $P_T^2=2\rho_T V^2$, where  $V$ is the volume at  $T$. 
At the bounce---when the energy density reaches the  value
 $\rho_{\rm crit}\approx  0.41\, \rho_{\text{Pl}}$ and the volume is minimized at $V_{\text{b}}$---we have:
\begin{equation}
P_T=V_{\rm b}\sqrt{2\rho_{\rm crit}}.
\end{equation}
Beyond the bounce, although $V$ grows rapidly, $P_T$ remains fixed by this  initial condition. In any realistic, semiclassical universe, $V_{\text{b}}$ must far exceed the Planck volume $V_{\rm Pl}\equiv\ell_{\text{Pl}}^3$. For a closed FLRW model, one typically assumes $V_{\text{b}} \sim 10^5V_{\rm Pl}$--$10^{20}V_{\rm Pl}$, depending on the bounce scale.%
\footnote{By contrast, in a spatially flat ($k=0$) setting, one often considers even larger volumes, up to $10^{100}V_{\rm Pl}$.} For instance, if  $V_{\text{b}} \sim 10^{8} V_{\rm Pl}$, then
\begin{equation}
P_T \approx 10^8\ell_{\text{Pl}}^3 \sqrt{0.82\rho_{\rm Pl}} \sim 9\times 10^7,
\end{equation}
and hence,
\begin{equation}
\langle \hat{H}_o^{-1}\rangle_o \sim  10^{-15}.
\end{equation}
If $V_{\text{b}}$  grows to $10^{10}V_{\rm Pl}$ (more realistic for our Universe), one finds
\begin{equation}
\langle \hat{H}_o^{-1}\rangle_o \sim 10^{-19}.
\end{equation}
In this  regime, a massive fermion's backreaction energy density [cf. Eq.~(\ref{eq:EnergyOperator-2})] approaches a constant and can be interpreted as an effective cosmological constant,
\begin{equation}
\Lambda_{\rm eff} = 8\pi G\rho_n = 8\pi G m \langle \hat{H}_o^{-1} \rangle_o,
\end{equation}
with $G=1$ in Planck units. Taking $m\sim 10^{-19}m_{\rm Pl}$ (roughly a proton mass) yields
\begin{equation}
\Lambda_{\rm eff} \sim 6 \times 10^{-33},
\end{equation}
whereas the observed value is $\Lambda_{\rm obs} \sim 10^{-122}$. Hence, even standard-mass fermions in a  $k=1$ LQC universe produce $\Lambda_{\text{eff}}$ roughly $10^{89}$ times larger than 
$\Lambda_{\rm obs}$. To reconcile this mechanism with late-time acceleration would therefore require either an ultralight fermion ($m\ll10^{-19}m_{\rm Pl}$) or an exceedingly large bounce volume---conditions well beyond those expected in a semiclassical $k=1$ cosmology.

\section{Conclusion and Outlook}
\label{sec:conclusion}

In this work we have studied the dynamics of a Dirac fermion field on a loop-quantized, closed FLRW background. 
The background geometry is coupled to a massless scalar field $T$, which serves as internal time.  By expanding the fermionic field in terms of spinor harmonics on $\mathbb{S}^3$, we derive the Hamiltonian governing its evolution and establish the following key results:
\begin{itemize}
\item[(1)]  In the test-field approximation---where the backreaction of fermionic modes on the background is neglected---each fermionic mode evolves independently on an emergent dressed metric. The components of this metric are determined by the quantum fluctuations of the underlying geometry. Notably, massive modes perceive distinct dressed metrics whose temporal and spatial components both carry Planck-scale corrections. Massless modes---being conformally coupled---do not sample a unique geometry; instead they feel a  family of conformally invariant backgrounds in which the spatial sector remains classical and only the temporal component is dressed by quantum effects.

\item[(2)] Going beyond the test-field approximation via a Born-Oppenheimer approximation, we allowed fermion backreaction to shift the background Hamiltonian constraint. Each mode's energy $E_n=\pm w_n$ [cf.~Eq.~(\ref{eq:operatorV-E})] (corresponding to the fermion's vacuum and pair states) induces a positive or negative contribution $\Theta_n(v)$ [cf.~Eq.~(\ref{eq:constraintBRTerm5-A})]  to the Hamiltonian of the quantum background.
As a result, each massive mode experiences its own distinct rainbow metric, while massless modes see  a conformally invariant class of rainbow metrics ($\mathbb{S}^3$ unchanged, time rescaled, and mode specific). 
\end{itemize} 
The backreaction effects differ fundamentally between fermionic and bosonic fields due to their distinct quantization properties. Fermionic modes are restricted to exactly two discrete energy eigenstates  [$E_n(v)=\pm w_n$], producing precisely two possible backreaction configurations for each mode. In contrast,  scalar (bosonic) fields support a countably infinite set of discrete energy states, generating  strictly positive spectra of backreaction possibilities that give rise to an infinite family of possible rainbow metrics \cite{Parvizi:2021ekr}. The underlying reason is the finite-dimensional  Hilbert space for fermions, versus the infinite tower of bosonic excitations seen by scalar fields.

Our findings have significant phenomenological implications across multiple cosmological regimes, from the deep Planck scale to late-time evolution, with potentially observable consequences. The vacuum occupation state of fermion modes [characterized by  $E_n=-w_n$, $\Theta_n(v)>0$] elevates the effective minisuperspace potential barrier, consequently delaying the quantum bounce to smaller volumes (higher-energy densities). Conversely, excited occupation states  [$E_n=+w_n$, $\Theta_n(v)<0$] reduce this barrier, causing the bounce to occur at larger volumes (lower-energy densities). This mode-dependent behavior may lead to a breaking of time-reversal symmetry, establishing a cosmological arrow of time through asymmetric pre- and postbounce evolution.
Notably, massive fermions develop an approximately constant energy density at large volumes, effectively mimicking an emergent cosmological constant, while the backreaction of massless fermions decays dynamically and cannot sustain cosmic acceleration. The discrete, mass-dependent rainbow metric corrections to the background spacetime geometry may produce distinctive signatures in several observational channels, including (i) modifications to primordial perturbation spectra, (ii) altered neutrino propagation characteristics, and (iii) potential imprints on late-time accelerated expansion.

Several important directions warrant further investigation to advance our understanding of these quantum gravitational effects: First, numerical simulations of the full difference-equation evolution for fermionic modes across the bounce are needed to quantitatively assess pair production and backreaction effects. Second, the inclusion of gauge interactions would provide a more complete description of early-Universe dynamics, particularly in the Planck epoch. Third, it would be valuable to explore whether alternative quantization approaches---such as the Hartle-Hawking boundary proposal \cite{Hartle:1983ai} or the D'Eath-Halliwell framework \cite{DEath:1984gmo}---applied to spinor fields might similarly generate a rainbow metric structure. These investigations would not only elucidate the interplay between spin, mass, and quantum geometry but could also reveal novel observational signatures of quantum gravity in the early Universe.

\section*{ACKNOWLEDGMENT}

The early stages of this work were inspired by insightful discussions with the late  Jurek Lewandowski, whose profound understanding of loop quantum gravity and cosmology and intellectual generosity greatly influenced the development of these ideas. It is with sincere gratitude that we dedicate this paper to his memory. He will be profoundly missed as both a remarkable scientist and a cherished  colleague and mentor.

Y.T. acknowledges support from the University of Warsaw within the IDUB program (Grant No. PSP: 501-D111-20-0023110) and from the Iranian Bonyad-e Melli-e Nokhbegan (BMN) via the Kazemi-Ashtiani grant. This work was also conducted within the framework of the following COST (European Cooperation in Science and Technology) Actions: CA18108 ``Quantum Gravity Phenomenology in the Multi-Messenger Approach,'' CA23130 ``Bridging High and Low Energies in Search of
Quantum Gravity,'' and CA23115 ``Relativistic Quantum Information.''

\section*{DATA	AVAILABILITY}

The data that support the findings of this study are available from the corresponding author upon reasonable request.

\appendix

\section{Solutions to the eigenvalue equation}
\label{sec:eigenvalue}

We aim to solve the eigenvalue equation (\ref{eq:Eigenvalue1}), where \( x, y \) are Grassmann-valued variables. These variables satisfy the anticommutation relations \( xy = -yx \) and \( x^2 = y^2 = 0 \). Given this structure, the most general form of the fermion's wave function $\psi_{np}$ is
\begin{equation}
\psi_{np}(x,y) = c_0 + c_1x + c_2y + c_3xy,
\label{eq:psiGrassmann}
\end{equation}
where \(c_0, c_1, c_2, c_3\) are ordinary (non-Grassmann) functions. Substituting  \(\psi_{np}(x,y)\) from Eq.~(\ref{eq:psiGrassmann}) into Eq.~(\ref{eq:Eigenvalue1}), we obtain
\begin{align}
&\lambda_n\ell^{-2}  V^{\frac{2}{3}}\left(-c_0   +c_3xy\right)
+  m\ell^{-3} V\left(c_3 - c_0xy\right) 
= E_{np}(c_0+c_1x+c_2y+c_3xy).
\end{align}
Since both sides must match term by term in the Grassmann expansion, we equate coefficients (of $1$, $x$, $y$, and $xy$, separately) and obtain the corresponding equations for the unknown variables $c_0, \ldots, c_3$. By solving these equations we obtain the unknown variables.

From the constant terms (corresponding to the unit term in Grassmann expansion),  we obtain:
\begin{equation}
\left(\lambda_n\ell^{-2}  V^{\frac{2}{3}}  +E_{np}\right)c_0 
= m\ell^{-3} V c_3.
\label{eq:c_0c_3}
\end{equation}
From the $xy$ terms, we get
\begin{equation}
\left(\lambda_n\ell^{-2}  V^{\frac{2}{3}} - E_{np}\right)c_3 = m\ell^{-3} Vc_0 .
\label{eq:c_3c_0}
\end{equation}
The linear term in \( x \)  yields
\begin{equation}
E_{np} c_1=0 \quad \Rightarrow \quad c_1 = 0 \quad \text{or} \quad E_{np}=0.
\label{eq:c_1}
\end{equation}
Similarly, for the linear term in \( y \)  we get
\begin{equation}
E_{np} c_2=0 \quad \Rightarrow \quad c_2 = 0 \quad \text{or} \quad E_{np}=0.
\label{eq:c_2}
\end{equation}
The nature of the solution depends on the choice of  $E_{np}$.  In the following, we will analyze the physically relevant solutions to above equations describing the fermionic  quantum states.

From Eqs.~(\ref{eq:c_0c_3}) and (\ref{eq:c_3c_0}), we get a linear system 
\begin{equation}
\begin{pmatrix}
\lambda_n\ell^{-2}  V^{\frac{2}{3}}  +E_{np} & -m\ell^{-3} V \\
-m\ell^{-3} V  & \lambda_n\ell^{-2}  V^{\frac{2}{3}}   - E_{np}
\end{pmatrix}
\begin{pmatrix}
c_0 \\
c_3
\end{pmatrix}=0.
\label{eq:linearsystem}
\end{equation}
For nontrivial solutions, the determinant must vanish. This yields two solutions for $E_{np}$,
\begin{equation}
E_{np}^{(\pm)} =\pm\sqrt{\lambda_n^2\ell^{-4}  V^{\frac{4}{3}}  + m^2\ell^{-6} V^2} \equiv \pm w_{n}.
\end{equation}
By setting these  two solutions for $E_{np}$ into the system of equations (\ref{eq:linearsystem}), we obtain two relations between $c_0$ and $c_3$ as
\begin{equation}
c_0 = \frac{m\ell^{-3} V}{\lambda_n\ell^{-2}  V^{\frac{2}{3}}  +E_{np}}c_3, \quad \text{with}\quad E_{np}^{(\pm)}=\pm w_n,
\end{equation}
or, equivalently,
\begin{align}
E_{np} &= +w_{n}: \quad \quad  c_3 = \frac{m\ell^{-3} V}{\lambda_n\ell^{-2}  V^{\frac{2}{3}}  - w_{n}}c_0, \\ 
E_{np} &=-w_{n}: \quad \quad    c_3= \frac{m\ell^{-3} V}{\lambda_n\ell^{-2}  V^{\frac{2}{3}} + w_{n}}c_0.
\end{align}
Consequently, the corresponding (two) eigenstates are superpositions of $1$ and $xy$,
\begin{equation}
\psi_{np}^{(\pm)} = c_0\left(1+\frac{m\ell^{-3} V}{\lambda_n\ell^{-2}  V^{\frac{2}{3}}  \mp w_{n}}xy\right).
\end{equation}
Likewise,  Eqs.~(\ref{eq:c_1}) and (\ref{eq:c_2}) yield  either $E_{np}=0$ with $c_1, c_2\neq 0$ or $E_{np}\neq0$ with $c_1=c_2=0$. If $E_{np}\neq0$, then \( c_1=c_2 = 0 \) leads to the trivial solution \( \psi(x,y) = 0 \), which is uninteresting. If \( E_{np} = 0 \), then Eq.~(\ref{eq:psiGrassmann}) simplifies to two different solutions,
\begin{align}
E_{np}=0: \quad \quad \psi_{np}= c_1x \quad \text{or} \quad 
\psi_{np}= c_2y.
\end{align}
Therefore, we obtained four solutions for the wave function of the Dirac spinor modes summarized in  Eqs.~(\ref{eq:EigenValueSoltogenTOT}).

\section{Difference operator in harmonic gauge}
\label{app:DifferenceOp}

The quantized gravitational Hamiltonian for the closed FLRW model can be expressed into two parts as follows [cf. Eq.~(\ref{eq:Hamiltonian-GravQ})]:
\begin{subequations}
\begin{align}
\Theta_0 \Psi_o(v)&= -\frac{1}{8\pi G \hbar^2}\left[\hat{V}^{\frac{1}{2}}e^{i f \ell_o} \sin(\bar{\mu} c) \hat{A} \sin(\bar{\mu} c) e^{-i f \ell_o}\hat{V}^{\frac{1}{2}}\right]\Psi_o(v),   \\
\Theta_1\Psi_o(v) &= \frac{1}{8\pi G\hbar^2}\hat{V}^{\frac{1}{2}}\left[\sin^2 \left( \frac{\bar{\mu}\ell_o}{2} \right) - \frac{\bar{\mu}^2\ell_o^2}{4} -\frac{\ell_o^2}{9 |K^2 v|^{2/3}} \right] \hat{A}\hat{V}^{\frac{1}{2}} \Psi_o(v).
\label{eq:Hamiltonian-GravQ-App}
\end{align}
\end{subequations}
The operation of sine and cosine terms on the background wave function are given by
\begin{align}
\sin(\bar{\mu}c) \Psi_o(v) &= \tfrac{1}{2i}\left(e^{i\bar{\mu}c} - e^{-i\bar{\mu}c}\right)\Psi_o(v) \nonumber \\
&= \tfrac{1}{2i}\left[\Psi_o(v+2) - \Psi_o(v-2)\right], \nonumber \\
\cos(\bar{\mu}c) \Psi_o(v) &= \tfrac{1}{2}\left(e^{i\bar{\mu}c} + e^{-i\bar{\mu}c}\right)\Psi_o(v) \nonumber \\
&= \tfrac{1}{2}\left[\Psi_o(v+2) + \Psi_o(v-2)\right].
\end{align}
Additionally, the operator \( \hat{A} (v) \) is  given by [cf. Eq.~(\ref{eq:A-hat})]:
\begin{align}
  \hat{A} \Psi_o(v) &=  C_A(v)\,  \Psi_o(v),
\end{align}
where $C_A(v)$ is a function of $v$ defined by
\begin{equation}
C_A(v) \equiv \frac{27K}{4} \sqrt{\frac{8\pi}{6}} \frac{\ell_{\rm Pl}}{\gamma^{3/2}}\, |v| \big||v+1| - |v-1| \big|.
\end{equation}
To  compute  $\Theta\Psi_o$, we successively apply each factor in $\Theta_0$ and $\Theta_1$. Below, we outline the computation for each term.

In term $\Theta_0\Psi_o$, the rightmost $\hat{V}^\frac{1}{2}$ acts on  $\Psi_o$ according to the eigenvalue equation (\ref{eq:volumeoperatorEV}) for the volume operator:
\begin{align}
\Psi_1(v)\equiv\hat{V}^\frac{1}{2}\Psi_o(v) = \sqrt{C_V(v)}\Psi_o(v), 
\end{align}
where $C_V(v)$ is the eigenvalue of the volume operator,
\begin{align}
C_V(v)\equiv \left(\frac{8\pi \gamma}{6}\right)^{\frac{3}{2}}\frac{\ell_{\rm Pl}^{3}}{K}|v|.
\end{align}
Next, applying $\sin(\bar{\mu}c)$ yields
\begin{align}
\Psi_2(v) &\equiv \sin(\bar{\mu}c) e^{-if\ell_o}\Psi_1(v) 
\nonumber \\
&= \frac{1}{2i}\sqrt{C_V(v)}\, e^{-if\ell_o}\left[\Psi_o(v+2)- \Psi_o(v-2)\right].
\end{align}
Acting $\hat{A}$ on $\Psi_2$ results in
\begin{align}
\hat{A}\Psi_2(v) &= \frac{1}{2i}\sqrt{C_V(v)}\, e^{-if\ell_o}\hat{A}\left[\Psi_o(v+2)- \Psi_o(v-2)\right] 
\nonumber\\
&= \frac{1}{2i}\sqrt{C_V(v)}\, e^{-if\ell_o}
 \Big[C_A(v+2) \Psi_o(v+2) 
  - C_A(v-2)\Psi_o(v-2) \Big].
\end{align}
Applying $\sin(\bar{\mu}c)$ again gives
\begin{align}
\Psi_3&\equiv e^{if\ell_o}\sin(\bar{\mu}c)\hat{A}\Psi_2(v)  \nonumber\\
&= -\frac{1}{4}\sqrt{C_V(v)}\,
 \Big[C_A(v+2) \big(\Psi_o(v+4) - \Psi_o(v)\big) \nonumber \\
 & \qquad \qquad   \qquad \quad  - C_A(v-2)\big(\Psi_o(v) - \Psi_o(v-4)\big)\Big].
\end{align}
Finally, applying  $\hat{V}^{1/2}$ on $\Psi_3$  from the left-hand side yields
\begin{align}
\Theta_0\Psi_o(v) &= 
-\frac{1}{8\pi G\hbar^2}\hat{V}^\frac{1}{2}\Psi_3(v)  \nonumber\\
&= \frac{1}{32\pi G\hbar^2}\sqrt{C_V(v)}\,
 \hat{V}^\frac{1}{2}\Big[C_A(v+2) \big(\Psi_o(v+4) - \Psi_o(v)\big) 
\nonumber \\
 & \qquad \qquad   \qquad \quad  - C_A(v-2)\big(\Psi_o(v) - \Psi_o(v-4)\big)\Big]
 \nonumber \\
&= \frac{1}{32\pi G\hbar^2}\sqrt{C_V(v)}\,
 \Big[C_A(v+2) \big(\sqrt{C_V(v+4)}\Psi_o(v+4) - \sqrt{C_V(v)}\Psi_o(v)\big) 
\nonumber \\
 & \qquad \qquad   \qquad \quad  - C_A(v-2)\big(\sqrt{C_V(v)}\Psi_o(v) - \sqrt{C_V(v-4)}\Psi_o(v-4)\big)\Big]
\nonumber\\
&= \frac{3\pi G}{8} 
 \Big[C^+(v)\Psi_o(v+4) +C^0(v) \Psi_o(v)\big) 
+C^-(v)\Psi_o(v-4)\Big],
\end{align}
where we have defined
\begin{align}
C^{+}(v) &= |v+2| \big||v+3| - |v+1| \big| \sqrt{|v||v+4|}, \nonumber \\
C^{0}(v) &= - \big(|v+2| \big||v+3| - |v+1| \big| + |v-2| \big||v-1| - |v-3| \big|\big)|v|,\nonumber  \\
C^{-}(v) &=  |v-2| \big||v-1| - |v-3|\big| \sqrt{|v||v-4|}.
\label{eq:Theta0a}
\end{align}
For positive volume, this reduces to Eq.~(\ref{eq:Theta0b}).
This corresponds to the first term in Eq.~(\ref{eq:Hamiltonian-GravQ-App}).

For $\Theta_1\Psi_o(v)$, we begin by computing
\begin{align}
\Psi_4(v) &\equiv \hat{A}\hat{V}^\frac{1}{2}\Psi_o(v) =\hat{A}\Psi_1 \nonumber \\
&= \sqrt{C_V(v)} C_A(v)\, \Psi_o(v).
\end{align}
Next, applying the terms in the bracket of $\Theta_1$  yields
\begin{align}
\Psi_5(v) &\equiv \left(\sin^2 \left( \frac{\bar{\mu}\ell_o}{2} \right) - \frac{\bar{\mu}^2\ell_o^2}{4} -\frac{\ell_o^2}{9 |K^2 v|^{2/3}} \right)\Psi_4(v) \nonumber \\
&=  \sqrt{C_V(v)}\,  C_A(v)\,  \left(\sin^2 \left( \frac{\bar{\mu}\ell_o}{2} \right) - \frac{\bar{\mu}^2\ell_o^2}{4} -\frac{\ell_o^2}{9 |K^2 v|^{2/3}} \right) \Psi_o(v).
\end{align}
Finally, applying $\hat{V}^\frac{1}{2}$ from the left-hand side gives
\begin{align}
\Theta_1\Psi_o(v) &= \frac{1}{8\pi G\hbar^2}\hat{V}^\frac{1}{2}\Psi_5(v) \nonumber \\
&= \frac{3\pi G}{2}|v|^2 \big||v+1| - |v-1| \big| \left(\sin^2 \left( \frac{\bar{\mu}\ell_o}{2} \right) - \frac{\bar{\mu}^2\ell_o^2}{4} -\frac{\ell_o^2}{9 |K^2 v|^{2/3}} \right) \Psi_o(v).
\label{eq:Theta1a}
\end{align} 
For $v>0$ this relation becomes (\ref{eq:Theta1b}). This completes the computation of $\Theta_1\Psi_o$.

\bibliography{Bibliography}

\end{document}